\documentclass[final,pdflatex,sn-nature]{sn-jnl}
\makeatletter
\@twosidefalse
\@mparswitchfalse
\makeatother
\usepackage{graphicx}%
\usepackage{multirow}%
\usepackage{amsmath,amssymb,amsfonts}%
\usepackage{amsthm}%
\usepackage{mathrsfs}%
\usepackage[title]{appendix}%
\usepackage{xcolor}%
\usepackage{textcomp}%
\usepackage{manyfoot}%
\usepackage{booktabs}%
\usepackage{algorithm}%
\usepackage{algorithmicx}%
\usepackage{algpseudocode}%
\usepackage{listings}%
\usepackage{siunitx}
\usepackage{comment}
\usepackage{siunitx}
\usepackage{mathrsfs}
\usepackage{lineno}
\usepackage{datetime}
\usepackage{amsmath}
\usepackage{textgreek}

\usepackage{hyperref}[hidelinks]

\AtBeginDocument{\switchlinenumbers}

\theoremstyle{thmstyleone}%
\theoremstyle{thmstyletwo}%

\theoremstyle{thmstylethree}%

\usepackage{geometry}
\begin{document}
\renewcommand\linenumberfont{\normalfont\small}

\title[Article Title]{Precise Measurement of the Absolute Sky Brightness at \qtyrange[range-phrase = \text{--}, range-units=single]{60}{350}{\mega\hertz}}

\author*[1]{\fnm{Luke} \sur{McKay}}\email{Luke.McKay@csiro.au}

\author[2]{\fnm{Ravi} \sur{Subrahmanyan}}\email{Ravi.Subrahmanyan@csiro.au}

\author[1]{\fnm{Aaron} \sur{Chippendale}}\email{Aaron.Chippendale@csiro.au}

\author[3]{\fnm{Pietro} \sur{Bolli}}\email{Pietro.Bolli@inaf.it}

\author[3,4]{\fnm{Georgios} \sur{Kyriakou}}\email{Georgios.Kyriakou@uniroma1.it}

\author[1]{\fnm{Alex} \sur{Dunning}}\email{Alex.Dunning@csiro.au}

\author[1]{\fnm{Ronald} \sur{Ekers}}\email{Ron.Ekers@csiro.au}

\affil[1]{\orgdiv{CSIRO Space and Astronomy}, \orgname{CSIRO}, \orgaddress{\street{26 Pembroke Road}, \city{Marsfield},  \state{NSW} \postcode{2122}, \country{Australia}}}

\affil[2]{\orgdiv{CSIRO Space and Astronomy}, \orgname{CSIRO}, \orgaddress{\street{26 Dick Perry Avenue}, \city{Kensington}, \state{WA} \postcode{6151}, \country{Australia}}}

\affil[3]{\orgdiv{Arcetri Astrophysical Observatory}, \orgname{National Institute for Astrophysics}, \orgaddress{\street{Largo E. Fermi 5}, \postcode{50125} \city{Firenze}, \country{Italy}}}

\affil[4]{\orgdiv{Department of Physics}, \orgname{University of Rome `La Sapienza'}, \orgaddress{\street{piazzale Aldo Moro 5}, \postcode{00185} \city{Rome}, \country{Italy}}}



\abstract{We present a precision measurement of radio sky brightness over \qtyrange[range-phrase = \text{--}, range-units=single]{60}{350}{\mega\hertz}.  Accurate knowledge of the radio sky over these frequencies is essential for modelling foregrounds in experiments targeting cosmic dawn and the epoch of reionization. Measurements below \qty{1}{\giga\hertz} are also needed to understand the Galactic cosmic ray electron spectrum, constrain nanojansky radio source populations and dark matter models, understand origins of the diffuse radio background, and improve calibration of long-wavelength radio telescopes.  Here we show that current all-sky maps and the 2016 Global Sky Model (GSM2016) require substantial corrections over \qtyrange[range-phrase = \text{--}, range-units=single]{60}{350}{\mega\hertz}.  GSM2016 has an offset exceeding \qty{100}{\kelvin} below \qty{100}{\mega\hertz} and must be scaled up by a factor increasing from 1.2 below \qty{200}{\mega\hertz} to 1.5 at \qty{350}{\mega\hertz}.  This significantly increases previously inferred excess radio background, motivating review of faint source populations and dark-matter decay models.  Our measurement used a new receiver architecture that self-calibrates its noise contribution and bandpass in situ while connected to an antenna. We used a single, accurately modelled, log-periodic SKALA4.1 antenna on a \qty{40}{\metre} diameter SKA-Low station ground mesh.  Sky models scaled to our measurements can set the absolute flux-density scale for SKA-Low and other low-frequency radio telescopes.}

\maketitle

\section{Main} \label{sec:intro}

We have made a new, precision measurement of the radio sky using a new receiver with advanced architecture \citep{AU2025901969} that brings modern microwave engineering metrology methods from the laboratory to the field.  Extended Data Fig.~\ref{fig:radiometer_concept} shows the novel architecture of the receiver that enables continuous and precise calibration in situ when connected to an antenna in the field. The receiver is connected to a single, wideband log-periodic antenna that is fairly frequency-independent.  We accurately modelled the antenna with modern electromagnetic simulation tools to characterise radiation patterns and performance. The result is a precision measurement of the brightness of the diffuse sky from \qtyrange{60}{350}{\mega\hertz} that improves the absolute calibration of previous all-sky maps and global sky models at long wavelengths. 

Accurately calibrated low-frequency maps and sky models are required to account for foregrounds when making radio measurements of the early Universe.  They are particularly important when measuring small departures from the Planck spectrum of the cosmic background radiation that trace astrophysical processes in the early Universe during cosmic dark ages, cosmic dawn, and the epoch of reionization. They are further motivated by their potential to refine astrophysical constraints on source populations, cosmic ray propagation models, and dark matter decay.  Additionally, accurate sky models serve as a stable reference for setting the flux-density scale of low-frequency radio telescopes, including SKA-Low which will be the world's largest low-frequency telescope array.

At frequencies below about \qty{1}{\giga\hertz}, the radio sky brightness is dominated by synchrotron radiation processes in our Galaxy and by the sum of emission from all extragalactic sources \cite{1970Natur.228..847C}.  Fitting theoretical models to radio sky brightness measurements constrains cosmic ray particle acceleration and propagation in the interstellar gas and magnetic fields \citep{2013MNRAS.436.2127O}, as well as the populations of extragalactic radio sources \citep{2008ApJ...682..223G}.  Precision in the measurement of the radio sky brightness thus constrains the cosmic ray electron spectrum and models for any unaccounted-for extragalactic component that might arise from dark matter annihilation or decay \citep{2011PhRvL.107A1302F} or cosmological nanojansky populations \citep{2012ApJ...758...23C}.

The production, propagation and acceleration of cosmic ray leptons in the interstellar medium leads to a spatial distribution in their energy spectra.  This manifests in a sky distribution of Galactic synchrotron emission with breaks associated with energy spectrum knees.  For example, the GALPROP model is fitted to cosmic ray data from the Gamma Ray Observatory CGRO instruments EGRET, COMPTEL and OSSE \citep{1998ApJ...493..694M} and all-sky maps at long radio wavelengths \citep{2013MNRAS.436.2127O}. Inversely, the radio brightness distribution, spectral breaks, and magnetic fields constrain the lepton energy spectrum.
 
Balloon-borne ARCADE 2 radiometer \citep{2011ApJ...730..138S} measurements of the absolute brightness of the radio sky were analysed to model the Galactic emission \citep{2011ApJ...734....4K}.  This led to a suggestion for an unaccounted-for excess beyond models for Galactic diffuse emission and known populations of extragalactic sources \citep{2011ApJ...734....5F, 2011ApJ...734....6S}.  However, inferences of residual signals that spark explanations based on exotic astrophysics or sources \citep{2011PhRvL.107A1302F, 2012ApJ...758...23C} may arise from fitting physical models to inaccurate radio sky brightness \citep{2013ApJ...776...42S}.  

All-sky maps of the diffuse radio sky brightness have been made at several discrete frequencies (see \citep{2017MNRAS.464.3486Z} and references therein; \citep{Landecker19780_GalacticMetreWave,1999A&AS..137....7R, Haslam1981_408Observations, 1982A&AS...47....1H, 1982A&AS...48..219R, 1986A&AS...63..205R, 1997A&AS..124..315A,  1999A&AS..140..145M, 2001A&A...376..861R, 10.1093/mnras/stx1136, 2018AJ....156...32E}).  Images at long wavelengths are almost always made with antenna arrays, and the absolute scale has errors associated with uncertainties in absolute calibration of receiver systematics, antenna models, and ground emission. The all-sky maps that are the key input to the current Global Sky Model \citep[GSM2016;][]{2017MNRAS.464.3486Z} between \qty{45}{\mega\hertz} and \qty{1.42}{\giga\hertz}, bracketing the frequencies of our measurements, were observed between 1965 and 1999 \citep{1999A&AS..137....7R, 1997A&AS..124..315A, 1999A&AS..140..145M, 1982A&AS...47....1H, 1982A&AS...48..219R, 1986A&AS...63..205R,2001A&A...376..861R} but the primary absolute flux calibration measurements they rely upon were all made in the 1960s \citep{1962SvA.....6..122B, 1962MNRAS.124...61P, 1966Natur.210.1318H, 1969SvA....13..223P}.  Some of the maps are not directly tied to an absolute calibration, but are scaled to overlapping sections of older maps that have, in turn, been tied to an absolute calibration. Unsurprisingly, subsequent measurements with improved calibrations and using radiometers with better control of systematics have often indicated substantial errors in original maps (see \citep{Monsalve2021, 2024arXiv240906770W, 2015ApJ...801..138P, 10.1093/mnras/stx1136} for examples).

The Baars et al. flux scale \citep{1977A&A....61...99B} has been used to calibrate the majority of long-wavelength (\qty{> 1}{\centi\metre}) publications in radio astronomy for the last 50 years \citep{Kellermann2009BaarsCommentary}.  It collated accurate absolute measurements published between 1960 and 1975 in the \qty{10}{\mega\hertz} to \qty{35}{\giga\hertz} window of the strongest discrete sources---Cassiopeia~A, Cygnus~A and Taurus~A--- using single dipoles or horn antennas with accurately known effective areas.  Spectra of a set of secondary calibrators were then derived from \qty{400}{\mega\hertz} upwards. 

An updated flux scale for low-frequency radio telescopes over \qtyrange{30}{300}{\mega\hertz} \citep{2012MNRAS.423L..30S} was derived in 2012 in response to the construction of the Low-Frequency Array (LOFAR) telescope \citep{LOFAR2013}.  The spectra of a sample of discrete sources down to \qty{10}{\mega\hertz} was derived in 1973 \citep{1973AJ.....78.1030R} from historical measurements to provide a scale robust to the secular decrease in the flux density of Cassiopeia~A below \qty{100}{\mega\hertz} \citep{1990MNRAS.243..637R}. Measurements of six strong 3CR sources were rescaled to the robust scale of ref. \citep{1973AJ.....78.1030R} to provide best-fit spectral models for flux calibration of LOFAR and other low-frequency telescopes in the Northern Hemisphere.

 The Karl G. Jansky Very Large Array (VLA)  was used in 2017 \citep{2017ApJS..230....7P} to extend the spectral models for secondary calibrators in the Baars et al. scale \citep{1977A&A....61...99B} down to about \qty{50}{\mega\hertz}.  The Jansky VLA precisely measured flux-density ratios to translate the scale from 3C405 (Cygnus~A) to 20 sources that are distributed between both hemispheres.  However, despite the improvements in measurement precision, the absolute scale accuracy continues to be that which was set historically with single antennas to form the original Baars et al. scale \citep{1977A&A....61...99B}.

The Owens Valley Radio Observatory Long Wavelength Array (OVRO-LWA) was used in 2018 to make high fidelity maps of the northern radio sky between \qty{36.5}{\mega\hertz} and \qty{73.2}{\mega\hertz} \citep{2018AJ....156...32E} with tenfold improvement in angular resolution over existing maps.  The authors highlight the need for calibrated total power radiometry to improve the flux scale of their maps that is tied to the Baars et al. \citep{1977A&A....61...99B} spectrum of Cygnus~A. 

Ours is one of a number of contemporary efforts to improve the flux scale of low-frequency radio maps of the diffuse radio sky. Ref.~\citep{Guzman2011_AllSkyRadiation} combined existing maps at \qty{45}{\mega\hertz} \citep{1997A&AS..124..315A, 1999A&AS..140..145M} and made zero-point corrections to the resulting all-sky \qty{45}{\mega\hertz} map and the existing \qty{408}{\mega\hertz} all-sky map \citep{Haslam1981_408Observations, 1982A&AS...47....1H} by comparing these maps at six directions on the sky to a wide variety of published measurements between \qty{1}{\mega\hertz} and \qty{1420}{\mega\hertz}. Ref.~\citep{Monsalve2021} used calibrated total power radiometry data from EDGES \citep{Bowman2018_AbsorptionProfile} to correct the scale and offset of the combined all-sky \qty{45}{\mega\hertz} map of ref.~\citep{Guzman2011_AllSkyRadiation} and the \qty{150}{\mega\hertz} map of ref.~\citep{Landecker19780_GalacticMetreWave}.  EDGES used radiometry data from three different radiometers with scaled copies of blade antennas optimised for \qtyrange{50}{100}{\mega\hertz} (Low-Band 2), \qtyrange{60}{150}{\mega\hertz} (Mid-Band) and \qtyrange{90}{190}{\mega\hertz} (High-Band).  A unique contribution of our work is that we cover a wider frequency range of \qtyrange{60}{350}{\mega\hertz} with a single radiometer that has a new receiver architecture that continuously calibrates, in situ, for receiver noise, bandpass, and for transfer functions of signal power via antenna to receiver.  Wide frequency coverage about \qty{200}{\mega\hertz} is important as the radio sky spectrum has a change in steepness in the vicinity of that frequency \citep{Bridle1967_SpectrumRadioBackground}.

Given the uncertainties in measuring the flux densities of discrete sources at long wavelengths, we propose that the diffuse radio sky, enabled by our accurate measurements, might serve as a stable primary flux-density calibrator for long-wavelength radio astronomy instead of discrete sources. The diffuse radio sky was recently used to calibrate the Auger Engineering Radio Array \citep[AERA;][]{2025JInst..20P2017A}{} that measures short radio pulses from high-energy cosmic ray air showers.  It was also used to measure the system noise temperature and aperture efficiency of individual antennas of the Australian Square Kilometre Array Pathfinder (ASKAP) Telescope \citep{aces5, 2016PASA...33...42M, Chippendale2016EuCAP}.

The calibration of a radio interferometer would be a two-step process: (1) absolute calibration  of element auto-correlations via observations of the diffuse radio sky; (2) interferometer correlation coefficient measurements of strong unresolved sources where element gains have been corrected from step (1).  Flux calibration may then be quickly recovered during standard operation by making a short correlation measurement on one of the unresolved sources used at step (2).

Our measurements of the diffuse sky span all but the bottom \qty{10}{\mega\hertz} of the SKA-Low band of \qtyrange{50}{350}{\mega\hertz}.  For decades to come, the SKA will be the largest radio telescope array in the world operating at these frequencies so we describe an example of flux calibration for SKA-Low. The scale of each station autocorrelation signal is first set by comparing measured station auto-correlations to simulated station auto-correlations using an accurate model of the diffuse radio sky and accurate simulations of station beam patterns.  The results are used to scale all station auto-correlation measurements to kelvin units.  Next, inter-station correlation coefficients are measured on strong sources that are unresolved on measurement baselines.  These measurements are used to transfer the station calibrations to source strength.  These strong unresolved sources may then form the flux calibrator list for SKA-Low interferometer observations. 

\section{Results} \label{sec:results}
\subsection{Correcting the Global Sky Model}\label{sec:correcting_gsm}

We measured antenna temperature spectra from \qtyrange{50}{350}{\mega\hertz} on 2024 October 23 from 10:05:40 UTC to 18:00:06 UTC.  The measurements spanned nearly \qty{8}{\hour} of local sidereal time (20:00:52 LST to 03:56:36 LST), providing sufficient variation in Galactic contribution (\qty{>50}{\percent}) to accurately constrain both offset and scale corrections to GSM2016. Extended Data Fig.~\ref{fig:sky_coverage} illustrates the sky coverage of our measurements, plotting the radio sky in equatorial coordinates with truncation (grey fill) where there is no coverage due to limited LST range or the location of the antenna in the Southern Hemisphere. Our measurements were made wholly during the night to avoid increased uncertainty due to active solar radiation that may vary substantially. 

\begin{figure}[tb]
\centering
\includegraphics[trim=900 0 500 0,clip,width=0.7\textwidth,angle=-90,origin=c]{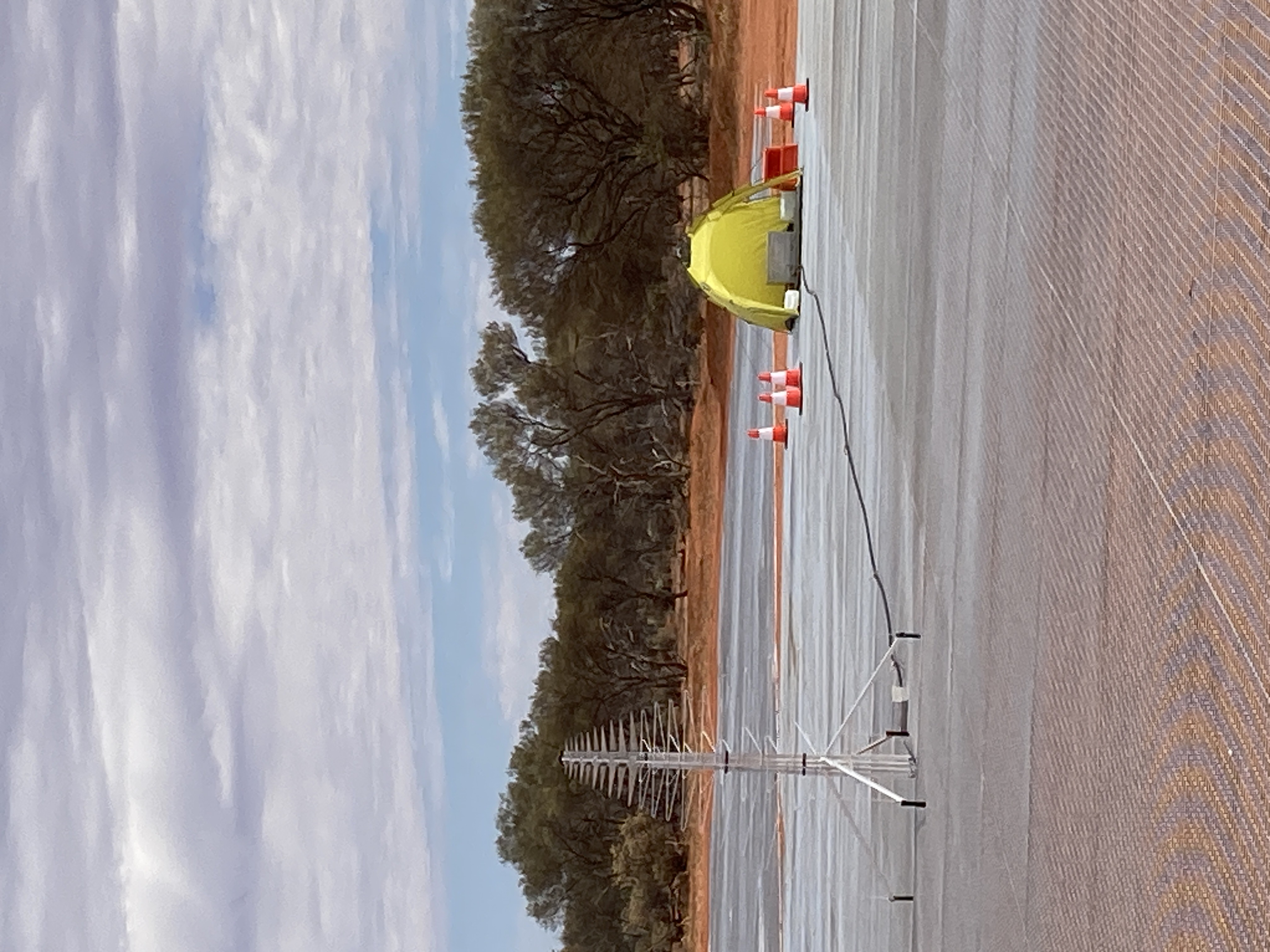}
\caption{\textbf{$|$ Experimental setup.} The observing system at Inyarrimanha Ilgari Bundara, the CSIRO Murchison Radio-astronomy Observatory in Australia. A single SKALA4.1 antenna \citep{9107113} at the centre of a \qty{40}{\metre} ground mesh is connected via a \qty{3.1}{\metre} coaxial cable to the smaller GINAN receiver box at the base of the antenna that contains an RF switch, calibration loads, and a noise source. The smaller box is connected via a \qty{19}{\metre} coaxial cable to the main GINAN receiver box in a tent at the edge of the mesh. See Extended Data Fig.~\ref{fig:receiver_small_box} for a close-up of the antenna and smaller box that is difficult to see here.}
\label{fig:Boolardy_Deployment}
\end{figure}

We used the new self-calibrating GINAN receiver \citep{AU2025901969} designed for precision absolute radiometry.  This was connected to a single SKALA4.1 antenna \citep{9107113} placed at the centre of a \qty{40}{m} diameter flat conductive ground mesh. Extended Data Fig.~\ref{fig:radiometer_concept} shows the novel receiver architecture and Fig.~\ref{fig:Boolardy_Deployment} shows a photo of the resulting radiometer system deployed for our measurement at Inyarrimanha Ilgari Bundara, the CSIRO Murchison Radio-astronomy Observatory in Australia.  We restrict analysis to \qty{60}{\mega\hertz} and above where the  SKALA4.1 antenna and mesh are best characterised with smooth frequency performance and negligible resistive losses.  Full receiver architecture, calibration procedure, antenna details, and uncertainty analysis are described in Methods.  

\begin{figure}[tb]
\centering
\includegraphics[trim=15 10 5 5,clip,width=1.0\textwidth,angle=0,origin=c]{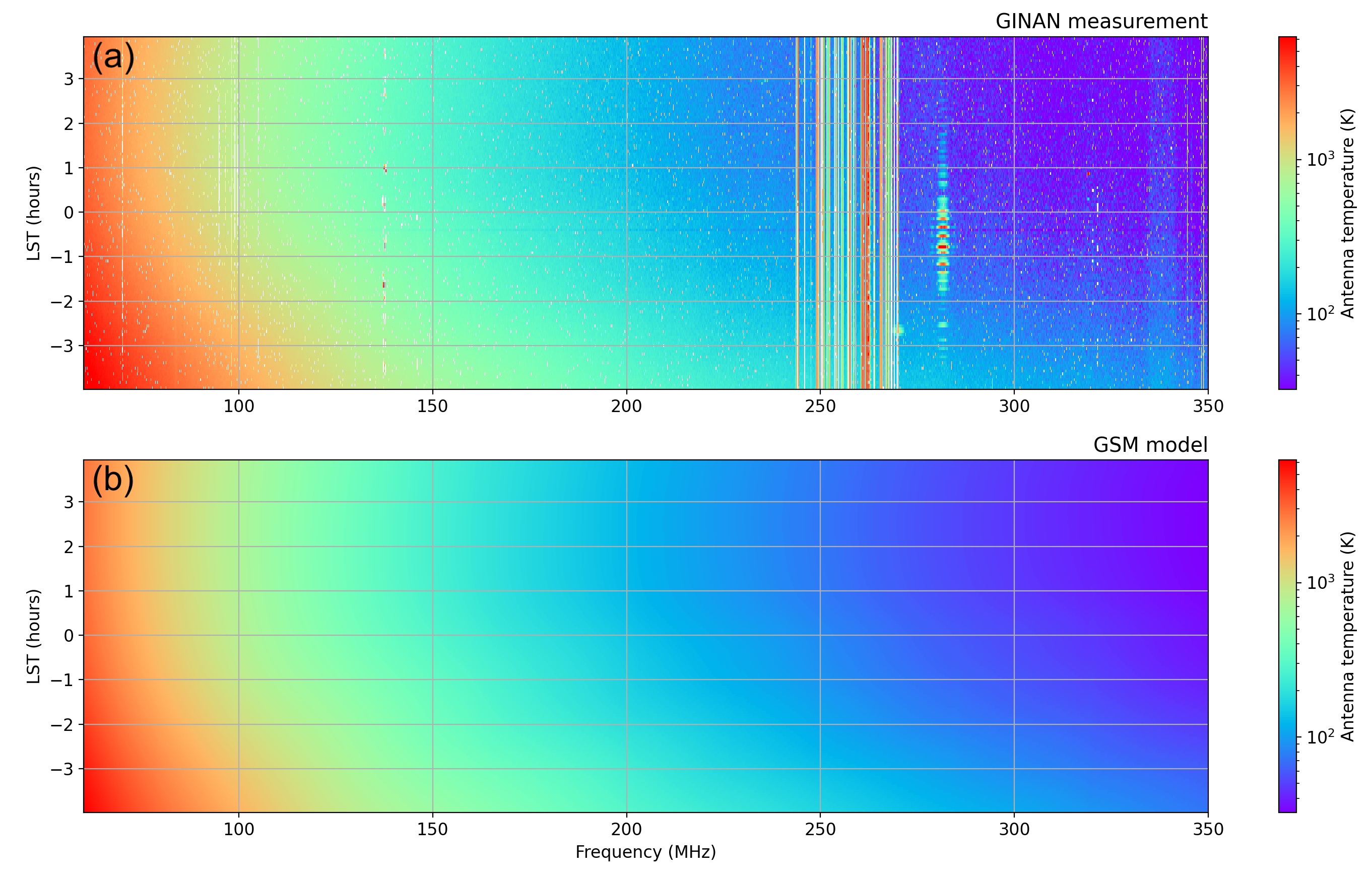}
\vspace{-3mm}
\caption{\textbf{$|$ Measured and modelled antenna temperatures.} Distribution of antenna temperature in time-frequency space, displayed in kelvin units. Colour bars at the right show the logarithmic scale that is common for both plots. \textbf{a}, Measured antenna temperatures. \textbf{b}, Predicted antenna temperatures from the Global Sky Model \citep[GSM2016;][]{2017MNRAS.464.3486Z} and simulated SKALA4.1 antenna pattern. 
\label{fig:TimeFreq_MeasModel}}
\vspace{2mm}
\centering
\includegraphics[trim=0 5 0 0,clip,width=1.0\textwidth,angle=0,origin=c]{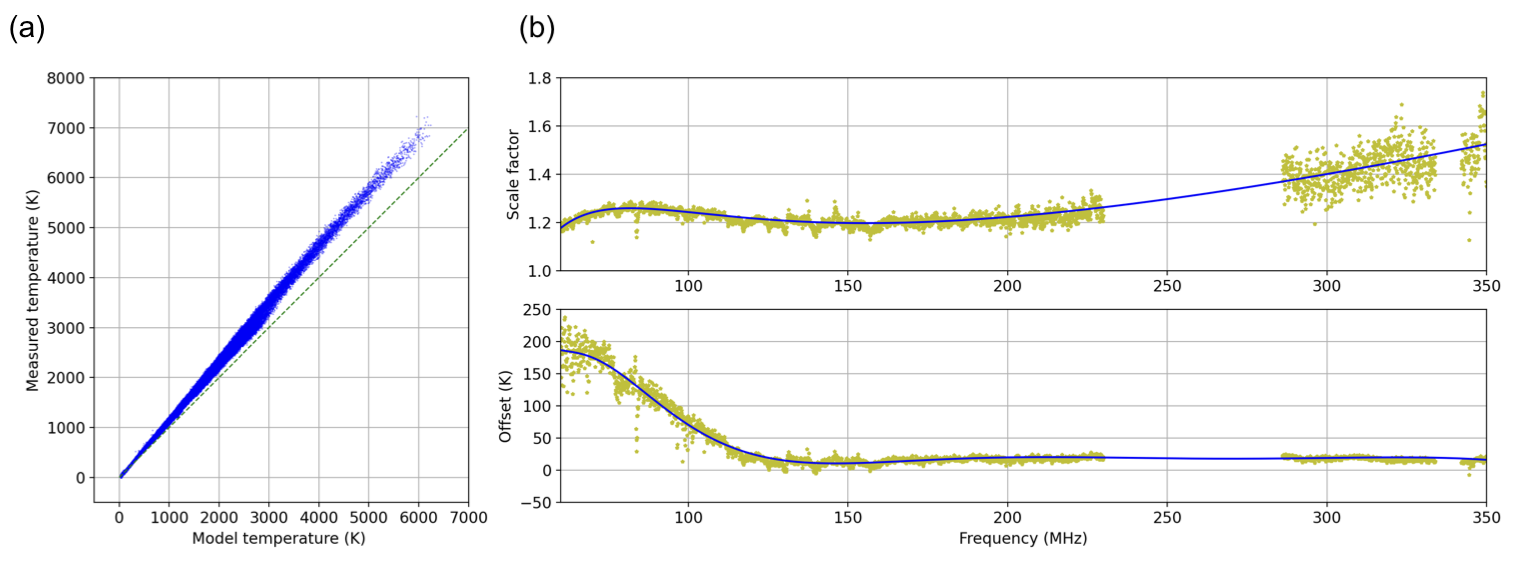}
\vspace{-2mm}
\caption{\textbf{$|$ Correcting the sky model.} \textbf{a}, Scatter plot of antenna temperatures showing GINAN measured temperature versus predicted temperatures calculated from the Global Sky Model \citep[GSM2016;][]{2017MNRAS.464.3486Z} and simulated SKALA4.1 antenna pattern. Blue markers overlay data for all frequencies and measurement epochs (\num{303830} data points).  All points would lie on the ideal dashed green line of unity slope and zero intercept if our measurements agreed with the GSM2016 prediction.  \textbf{b}, Derived corrections for model brightness temperatures; GSM2016 must have the offset subtracted and the remainder scaled up by the scale factor to match our measurements.  Olive markers show corrections solved for independently at each frequency to best fit data for all measurement epochs.  Solid blue lines show polynomial fits to the corrections given by equations~\eqref{eq:offset} and \eqref{eq:scale}.
\label{fig:TT_GSMcorrections}}
\end{figure}

Predictions for the antenna temperatures are derived using the improved Global Sky Model \citep[GSM2016;][]{2017MNRAS.464.3486Z} that refines the original model \citep[GSM2008;][]{2008MNRAS.388..247D} with new data sets and better interpolation algorithms.  Antenna patterns of an isolated SKALA4.1 antenna, placed on a \qty{40}{\metre} diameter conducting surface, have been computed over all azimuth and elevation angles from electromagnetic simulations using \texttt{FEKO} software.  The antenna measures dual linear polarization and the patterns have been computed for the single polarization mode used in the sky measurement and in \qty{1}{\mega\hertz} intervals over the frequency range of the observations.  At each local sidereal time in which sky data are acquired by the receiver, the radio sky at the observing site is computed from GSM2016 and the antenna temperature spectrum is predicted using the frequency-dependent antenna patterns. The antenna pattern is defined over elevation range \qtyrange{0}{90}{\degree} and over all azimuth angles.  Elevations below the horizon were not included when convolving antenna patterns with sky maps to predict antenna temperatures. Measurements were made with \qty{4}{\minute} cadence and predictions were computed with the same cadence and at the same sidereal times as measurements.

Fig.~\ref{fig:TimeFreq_MeasModel} displays the measurement and model predictions for the antenna temperatures, versus frequency and local sidereal time.  Despite the remote measurement site with relatively low radio frequency interference (RFI), measurement data still show some RFI.  It is persistent in some frequency channels and occasional in others. Data affected by RFI were masked and rejected from further processing. First, frequency channels with persistent RFI were masked, then Hampel filtering was performed in time and frequency domains to reject additional time-frequency pixels contaminated by RFI.  The complete band from \qtyrange{243.5}{270.4}{\mega\hertz} was rejected due to ultra-high frequency (UHF) satellite communications downlink emissions, along with bands \qtyrange{277.5}{285.0}{\mega\hertz} and \qtyrange{334}{342}{\mega\hertz} due to persistent RFI.

\begin{figure}[tb]
\centering
\includegraphics[trim=20 10 25 10,clip,width=1\textwidth,angle=0,origin=c]{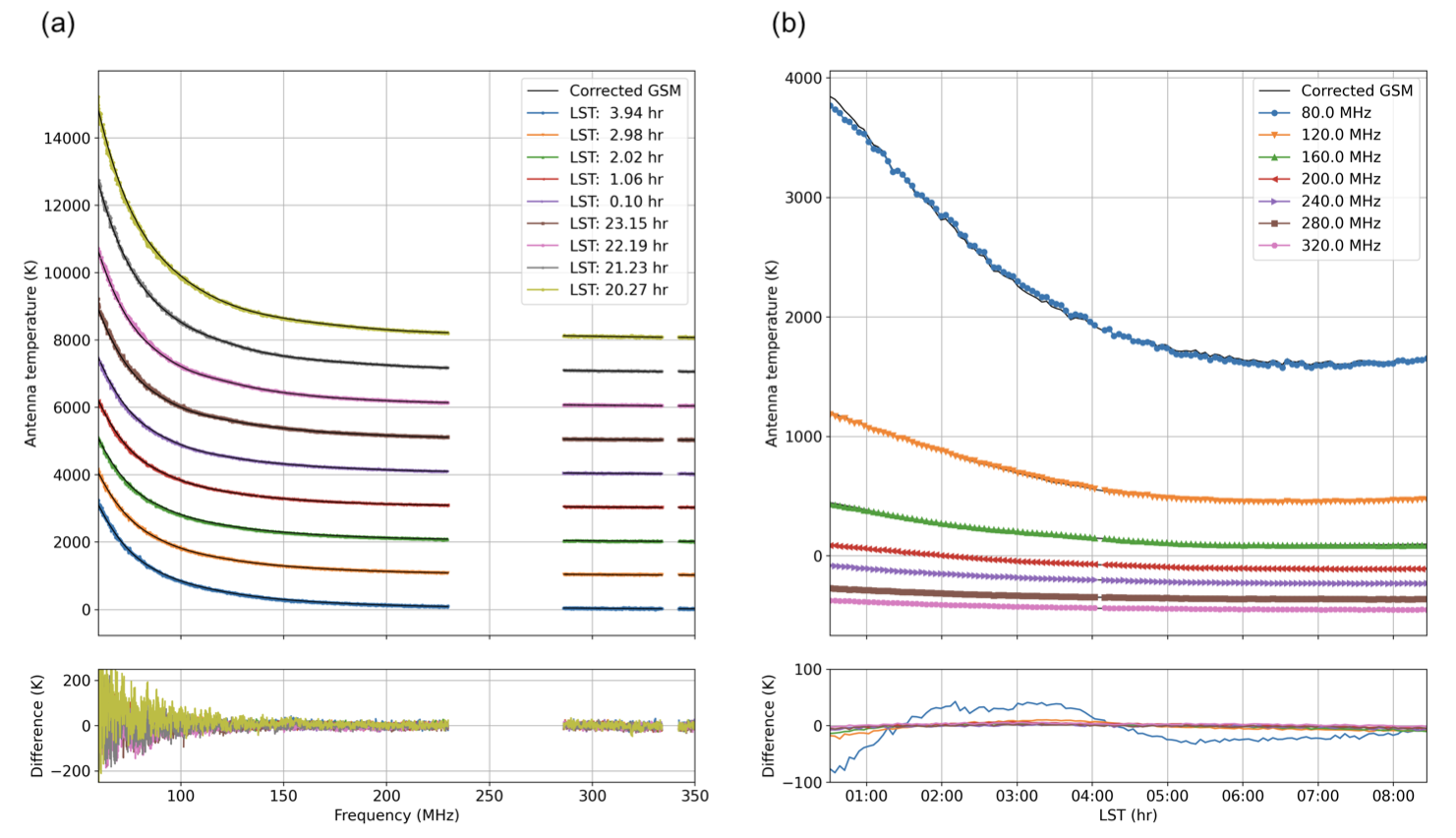}
\caption{\textbf{$|$ Measured vs. predicted antenna temperatures.} Upper panels show measured antenna temperatures (coloured markers) overlaid with predictions made using our corrected GSM (black traces).  Lower panels show the differences between measured and predicted temperatures.  Different LST times and frequencies are colour coded. 
\textbf{a}, Traces of antenna temperature versus frequency. Only every $15^\text{th}$ spectrum acquired over time is plotted, and successive spectra at times indicated in the legend have been offset vertically by \qty{1000}{\kelvin} for clarity. 
\textbf{b}, Antenna temperature versus local sidereal time. The data have been boxcar-averaged over \qty{40}{\mega\hertz} frequency bands and plotted at frequencies indicated in the legend. 
Successive traces at frequency \( f_\text{MHz} \) are displaced downward by \( 2(f_\text{MHz} - 70) \,\si{\kelvin} \). \label{fig:Meas_CorrGSM_TraceOverlays}}
\end{figure}

The distribution of antenna temperatures in a temperature-temperature plot of measured temperature versus model temperature is shown in Fig.~\ref{fig:TT_GSMcorrections}a, for all frequencies and times together (\num{303830} data points).  Within the uncertainty of the scatter, the model and measurements are in very close agreement and differ primarily by a scale factor.  However, the model best fits our data by adopting a two parameter correction that includes an offset as well as a scale factor and allows both parameters to vary with frequency. Fig.~\ref{fig:TT_GSMcorrections}b gives our resulting scale and offset corrections to GSM2016 that make it best fit our data.  We find that more than \qty{100}{K} must be subtracted from GSM2016 below about \qty{100}{\mega\hertz} and that, following this subtraction, the residual must be scaled up by a factor of approximately 1.2 below \qty{200}{\mega\hertz} rising to a factor of 1.5 at \qty{350}{\mega\hertz}. 

Our correction is summarised by two polynomials
\begin{equation}\label{eq:offset}
\begin{split}
T_{\rm offset}(f_{145})  =
\, \big[10.0086158 &- 9.03925211 \log f_{145} + 1523.27697 (\log f_{145})^2 -7270.93946 (\log f_{145})^3 \\
& - 7492.34793 (\log f_{145})^4 +68896.0788 (\log f_{145})^5 +8619.02374 (\log f_{145})^6 \\
&-198752.637(\log f_{145})^7 \big] ~ \unit{\kelvin}
\end{split} \end{equation}
\noindent {\rm and}
\begin{equation}\label{eq:scale}
\begin{split}
F_{\rm scale}(f_{145}) =
1.19851 &- 0.108045 \log f_{145} + 1.65211(\log f_{145})^2 + 2.86844(\log f_{145})^3 \\
&- 4.04712(\log f_{145})^4 + 7.41508(\log f_{145})^5
\end{split}
\end{equation}

\noindent where $f_{145}$ is the frequency in megahertz divided by 145, logarithms are base ten, and the correction is only valid from \qtyrange{60}{350}{\mega\hertz}. To best fit the measurement data, GSM2016 must first have $T_{\rm offset}$ subtracted, then the residual must be multiplied by $F_{\rm scale}$.

\begin{figure}[t]
\centering
\includegraphics[trim=70 0 90 0,clip,width=0.8\textwidth,angle=0,origin=c]{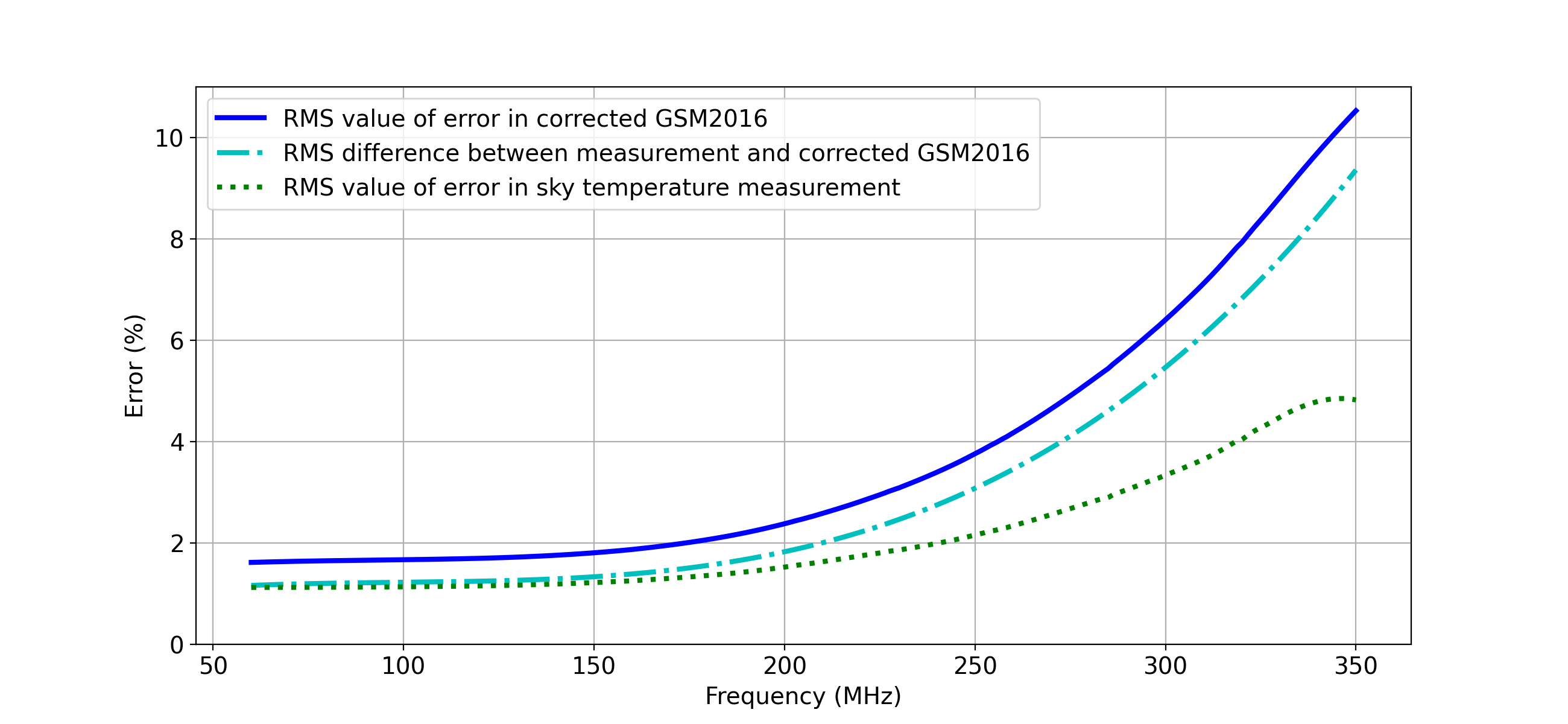}
\caption{\textbf{$|$ RMS values of errors}. 
The RMS error in our measurement of radio sky brightness temperature, versus frequency, is shown as a dotted line (see Section~\ref{sec:uncertainty}). The RMS value of the difference between our measured spectra and corrected GSM2016 predictions is shown as a dash-dot line.  This was evaluated over 24:00 LST to 04:00 LST when data amplitudes were reasonably constant because the Galaxy had set. These are combined in quadrature to yield the estimate of RMS error in corrected GSM2016 that is shown as a solid line.  All errors are provided as a percentage of the total sky brightness.
\label{fig:RMS_errors}}
\end{figure}

After correcting GSM2016, we compare it to our measurement data to estimate residual errors after the correction.  Fig.~\ref{fig:Meas_CorrGSM_TraceOverlays} shows measured antenna temperatures (coloured markers) overlaid with predictions for antenna temperatures using our corrected GSM2016 (black traces).  Sample antenna temperature spectra distributed over the range of local sidereal time of the observations are shown in Fig.~\ref{fig:Meas_CorrGSM_TraceOverlays}a and predictions for the antenna temperatures, averaged in \qty{40}{\mega\hertz} bins in frequency, are plotted versus local sidereal time in Fig.~\ref{fig:Meas_CorrGSM_TraceOverlays}b.  Fig.~\ref{fig:RMS_errors} shows the RMS value of the fractional difference between our corrected GSM2016 and the measurement data (dash-dot line) along with systematic uncertainties in our measurements (dotted line) that are detailed in Section \ref{sec:uncertainty}.  These are combined in quadrature to yield total RMS error in our corrected GSM2016 (solid line).

\subsection{Discussion}

We allowed our corrections to GSM2016 to be of two frequency-dependent parameters: an offset and a scale factor.  These were solved for in a least-squares fit of the model to the measurement data.  Following application of the best-fit correction, we show in Fig.~\ref{fig:RMS_errors} (dash-dot line) the RMS value of the fractional difference between our corrected GSM and the measurement data.  We calculated the RMS for the reduced range of 24:00 LST to 04:00 LST when data amplitudes were reasonably constant because the Galaxy had set.    Our corrected GSM deviates from measurements by less than \qty{2}{\percent} below \qty{200}{\mega\hertz}; however, this discordance increases with increasing frequency to about \qty{5}{\percent} at \qty{300}{\mega\hertz}.  The error in our corrected GSM appears dominated by the quality of the sky maps GSM2016 is built from and the algorithm it uses to interpolate between them to arbitrary frequency and sky direction.  The calibration error of our new measurements is significantly smaller as discussed in Methods (Sections~\ref{sec:calib_scale} and \ref{sec:uncertainty_abs_cal}).  We believe that the typical \qty{2}{\percent} error seen in Fig.~\ref{fig:RMS_errors} is inherent in GSM2016 and the legacy sky maps that it is built from.  The increase in error above \qty{250}{\mega\hertz} is likely due to the increasing impact of RFI in our data above that frequency (see Fig.~\ref{fig:TimeFreq_MeasModel}a).

\begin{figure}[b]
\centering
\includegraphics[trim=55 0 78 0,clip,width=0.8\textwidth,angle=0,origin=c]{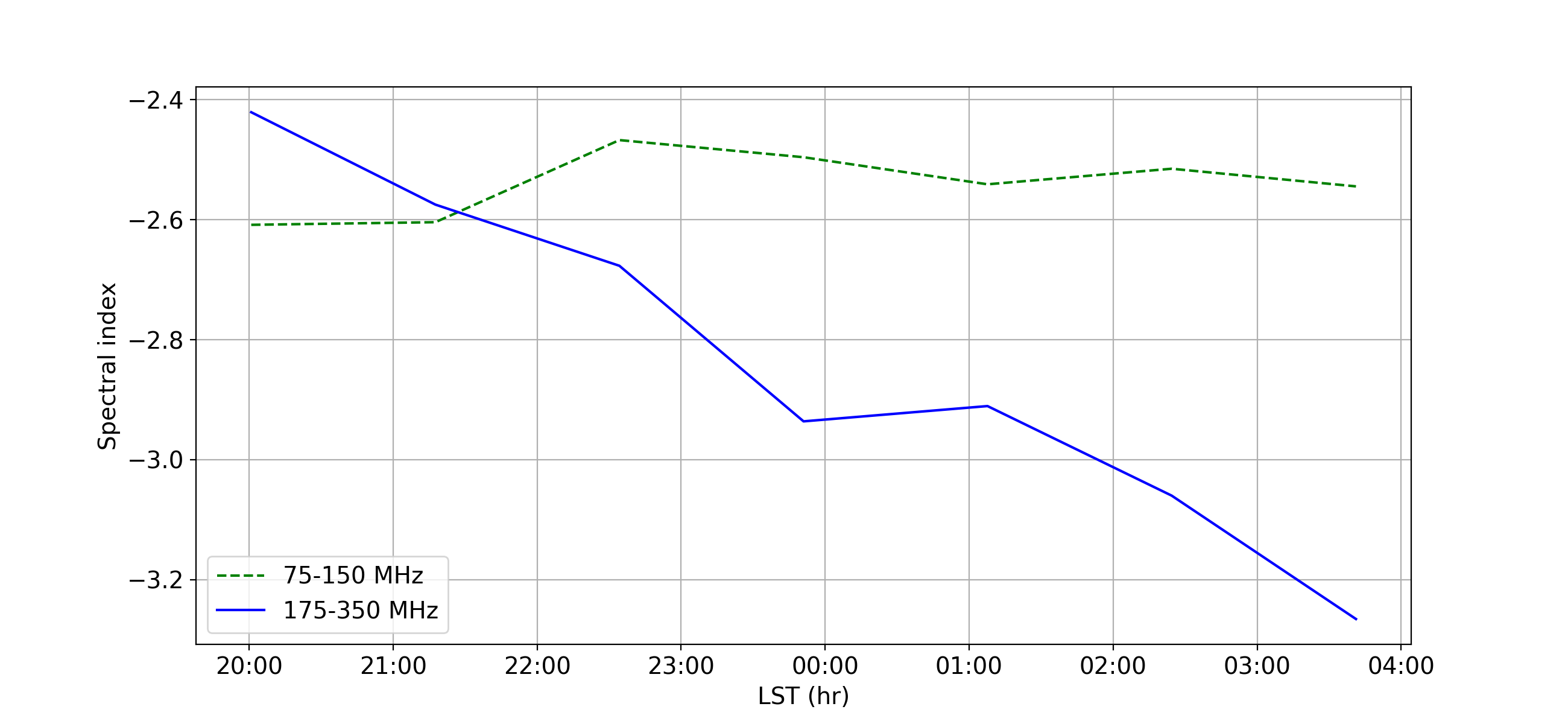}
\caption{The temperature spectral index of the radio sky measured by the GINAN receiver and SKALA4.1 antenna. Spectral index measured in the octave bands \qtyrange{75}{150}{\mega\hertz} and \qtyrange{175}{350}{\mega\hertz} are shown versus local sidereal time.
\label{fig:Spectral_index}}
\end{figure}

The sky spectrum, as measured using the SKALA4.1 antenna and the GINAN receiver, has a spectral index that varies over the band and over local sidereal time.  In Fig.~\ref{fig:Spectral_index} we show the variation in spectral index measured over the observing local sidereal time of 20:00~LST to 04:00~LST.  The temperature spectral index measured over \qtyrange{75}{150}{\mega\hertz} appears constant, with value about \num{-2.5}. Over the \qtyrange{175}{350}{\mega\hertz} band the index appears to steepen away from the Galactic plane, from about \num{-2.5} on the plane to as steep as \num{-3.2} off the plane. The radio sky spectrum in the \qtyrange{75}{350}{\mega\hertz} band thus has a break and is significantly more curved away from the Galactic plane. Our corrected GSM has spectral index behaviour in the \qtyrange{75}{350}{\mega\hertz} range that matches the measurement. 

EDGES radiometer measurements of spectral index in their low band \citep{Mozden2018_SpectralIndex} and high bands \citep{Mozden2017_ImprovedMeasurement} were made from the same latitude as ours.  They measured spectral index over the \qtyrange{50}{100}{\mega\hertz} band and, separately, over the \qtyrange{90}{190}{\mega\hertz} band.  This significantly overlaps our estimates for the \qtyrange{75}{150}{\mega\hertz} and \qtyrange{175}{350}{\mega\hertz} windows, although with different antenna patterns.  In the \qtyrange{0}{12}{\hour} LST range, which excludes the times when the Galactic centre transits, the EDGES measurement of temperature spectral index in the \qtyrange{50}{100}{\mega\hertz} band is in the range \numrange{-2.54}{-2.59}, consistent with our measurements.  In the \qtyrange{90}{190}{\mega\hertz} band, their measurement of spectral index in the same LST range is somewhat steeper and in the range \numrange{-2.60}{-2.62}, consistent with our inference that away from the Galactic plane the spectrum steepens towards higher frequencies.  When the wide antenna beam includes the Galactic centre, the EDGES data is interpreted to suggest a flattening of the spectral index to about \num{-2.50} in the \qtyrange{90}{190}{\mega\hertz} band, again consistent with our inference that the spectrum is flat over the \qtyrange{75}{150}{\mega\hertz} and \qtyrange{175}{350}{\mega\hertz} windows with index about \num{-2.5} as the Galactic plane dominates the radiometer beam solid angle. 

Spectral index measurements with the LEDA radiometers \citep{10.1093/mnras/stab1363} were made with a different antenna beam, at a different and northern latitude of \ang{+37.2398} and, with good instrument stability, over a different LST range of \qtyrange{9}{12.5}{\hour}.  Thus, their sampling is of a different region of the sky.  Their spectral index is determined over the band \qtyrange{50}{87}{\mega\hertz}, somewhat lower than our measurement in the \qtyrange{75}{150}{\mega\hertz} band.  Nevertheless, our measurements indicate that the sky spectrum is fairly uniform in index at these frequencies with a temperature spectral index in the range \numrange{-2.5}{-2.6}, which is consistent with their inference that the index is in the range \numrange{-2.50}{-2.56}. 

When we convolve the \qty{150}{\mega\hertz} all-sky survey map \citep{Landecker19780_GalacticMetreWave} with the radiation pattern of the SKALA4.1 antenna that we have used in our measurements, and compare with our data at \qty{150}{\mega\hertz}, we infer corrections in the form of a scale factor of \num{1.144(0.021)} and offset \qty{-3.53}{\kelvin}. An earlier comparison of the same map to EDGES data yielded a correction scale factor of \num{1.112(.023)} and correction offset \qty{0.7(6)}{\kelvin} \citep{Monsalve2021}.  The offset corrections are less than \qty{1}{\percent} of the data values and, therefore the significant correction is the scale factor.  Our scale correction is consistent with the scale correction derived from EDGES data \citep{Monsalve2021} to within quoted uncertainties. The small difference may be due to systematic errors in the absolute calibrations of our radiometer or the EDGES radiometers, or the limited LST range of our data compared to that of the EDGES data that had near complete LST coverage. 

OVRO-LWA measurements of the low-frequency sky have been set to the Baars et al. scale using Cygnus~A and secondary calibrators, to derive a low-frequency sky model with about \qty{20}{\percent} uncertainty in absolute flux-density scale \citep{10.1093/mnras/stx1136}.  Comparing with the original GSM2008, the OVRO-LWA measurements set to the Baars et al. scale  appear to have about a \qty{15}{\percent} higher temperature. The absolute accuracy of our measurements presented herein is substantially better than the OVRO-LWA results; nevertheless, within their errors, there is consistency between the OVRO-LWA results and our measurements. 

In this work we have accurately measured the low-frequency radio sky from \qtyrange{60}{350}{\mega\hertz} using the GINAN receiver and a log-periodic dipole SKALA4.1 antenna on a \qty{40}{m} ground mesh. These precision measurements constitute a first demonstration of the advanced GINAN receiver with a new architecture that provides accurate, dynamic and in situ characterisation of the receiver system bandpass, noise waves, and also the power transfer function from antenna to receiver including correction for impedance mismatch.  The key result of this work is the finding that the current Global Sky Model \citep[GSM2016; ][]{2017MNRAS.464.3486Z}, and hence the long-wavelength radio images that are the basis for its construction, require a significant frequency-dependent upward correction exceeding \qty{20}{\percent} to their absolute brightness. We have fitted for and provided a two-term correction factor for GSM2016 as a function of frequency.  Our primary data are made available \citep{McKay2026_RadioSkyBrightness} for setting the absolute flux-density scale of low-frequency telescopes including  SKA-Low.

Our results also impact interpretation of cosmic background radiation (CMB) and astrophysical cosmology that leaves imprints on it.  Global redshifted \qty{21}{\centi\metre} distortions in the CMB arise from galaxy-gas evolution across cosmic dawn and the epoch of reionization \citep{2012RPPh...75h6901P, 2019cosm.book.....M}.  Detecting these faint distortion signals requires careful modelling and separation of the foreground radio spectrum that is several orders of magnitude brighter.  Our precision measurement covers the radio spectral window in which these global \qty{21}{\centi\metre} CMB distortions are expected, and hence provides an accurate basis for the foreground separation.

The ARCADE 2 measurements of the absolute sky brightness, along with other data sets, were jointly analysed \citep{2011ApJ...734....5F, 2011ApJ...734....6S} to examine for any unaccounted-for excess. After fitting to the data and removing the Galactic emission, the CMB, and the integrated brightness from extragalactic sources, a uniform residual excess is inferred that has temperature spectral index $-2.62 \pm 0.04$ and brightness about \qty{171}{\kelvin} at \qty{150}{\mega\hertz} according to the recent revision of \citep{2025A&A...698A.152C} based on earlier work \citep{2011ApJ...734....6S, 2018ApJ...858L...9D}.  GSM2016 has mean sky brightness of \qty{392}{\kelvin} at this frequency of which known populations of extragalactic sources, assuming source counts given in \citep{2008ApJ...682..223G}, contribute about \qty{39}{\kelvin}. Our new absolute radio sky measurements require this unaccounted-for excess brightness $T_\text{excess}$ to be revised upwards to approximately \qty{201}{\kelvin} at \qty{150}{\mega\hertz}; an increase by a factor of \qty{17}{\percent}.  This yields a new estimate for the unaccounted-for uniform excess sky brightness of
\begin{equation}\label{eq:Texcess}
T_\text{excess} = (201 \pm 24)(f_\text{MHz}/\qty{150}{\mega\hertz})^{-2.62 \pm 0.04} \, {\unit{\kelvin}}.
\end{equation}

Explanations for the excess in terms of new source populations and/or dark matter decay now require the processes to have substantially greater radio photon production efficiency.  Having discussed this impact of our measurement on the excess, it may, however, be noted that this inferred excess is based on the specific method of Galactic modelling adopted in \citep{2011ApJ...734....5F}. This simplistic approach may have errors as pointed out by ref.~\citep{2013ApJ...776...42S} and more realistic modelling might mitigate the case for any significant excess.

\section{Methods}

\subsection{GINAN receiver with advanced architecture}\label{radiometer}

The GINAN receiver implements a new architecture that can dynamically self-calibrate for receiver noise and bandpass in situ, while connected to an antenna.  The receiver is self-contained in a portable shielded enclosure and designed to be connected directly to the terminals of a suitable antenna to make wideband absolute measurements of sky brightness. The receiver was primarily designed for detection of the global \qty{21}{\centi\metre} signal \citep{2012RPPh...75h6901P, 2019cosm.book.....M} from cosmic dawn and reionization, and has been repurposed in this work for absolute measurements of the diffuse radio sky. GINAN is a collaboration to make radio measurements of Global Imprints from Nascent Atoms to Now led by CSIRO.  Ginan is the name the Wardaman people of northern Australia give to the fifth brightest star of the Southern Cross which they saw as ``a small dilly bag full of knowledge, songs of knowledge that were passed on'' \citep{abcSouthernCross, DarkSparklers}.

Extended Data Fig.~\ref{fig:radiometer_concept} shows the radiometer receiver architecture that is protected by Australian patent application number 2025901969 filed on May 21, 2025 \citep{AU2025901969}. It is designed and constructed to dynamically measure the power transfer function through the antenna to the receiver, the receiver bandpass function, and the sum total of internally generated receiver noise from all parts of the receiver system, including amplifiers and lossy resistive elements, in terms of noise wave parameters \citep{1130490}. These are determined in situ, on a per-channel basis over the entire band, and repeatedly in intervals of minutes. They are used to correct the measured data to remove time-varying multiplicative transfer function and internally generated receiver noise. This provides a genuine precision spectrum of the radio sky over local sidereal time, without the need to stabilise the temperature of the receiver or manually calibrate it during observations. The data are recorded throughout the experiment and saved along with the temperatures of the terminations. Calibration corrections are calculated and applied during post processing. 

The receiver architecture uses both a vector network analyser (VNA) and a spectrometer. The RF switch connects the receiver input to one of five RF loads or the antenna. Measurements are taken in succession by both the VNA and the spectrometer before switching to the next load. After all five loads and the antenna have been measured, the process repeats throughout the experiment with a cadence of \qty{4}{\minute}. Each of the loads and the antenna is measured for an equal duration. During the VNA measurement, the output swept-frequency tone is generated at port 1 of the VNA. It passes through the coupled port of the directional coupler, through the switch, and then reflects off the load before continuing along the main propagation path in the receiver. The tone then returns to port 2 of the VNA through an output of the power splitter. The directional coupler ensures that all input loads are measured by the VNA and spectrometer without any change in system configuration. Additionally, the VNA has attenuators directly connected to both ports in order to increase isolation. These enable precise estimates of impedances, transfer functions, receiver noise, and power spectra. Established microwave engineering measurement methods are adopted to determine the electrical properties \citep{9116644, 10642982}.

The receiver is self-contained for remote deployment in the field, with a single-board computer for control and monitor, probes to sense physical temperatures of termination standards, solid-state drives to store data, and batteries to power the system. The receiver is housed inside an RF shielded enclosure to prevent RFI from the computer and VNA from coupling into the antenna. A paper detailing the receiver design and experimental results of laboratory qualification tests is in preparation.

\subsection{Observing system: GINAN receiver with SKALA4.1 antenna} \label{sec:observing_system}

Our new sky measurements were made with the GINAN receiver connected to a single SKALA4.1 antenna installed on a \qty{40}{\metre} diameter SKA-Low station ground mesh.  The resulting observing system is shown schematically in Extended Data Fig.~\ref{fig:radiometer_concept} and photographically in Fig.~\ref{fig:Boolardy_Deployment}.  Together, the novel receiver and the accurately characterised antenna enable an accurate measurement of the radio sky in the SKA-Low band.

SKALA4.1 is a dual linearly polarized log-periodic antenna, designed for SKA-Low with the goal of smooth frequency behaviour over the \qtyrange{50}{350}{\mega\hertz} band \citep{9107113}. However, below \qty{60}{\mega\hertz}, the antenna radiation efficiency  has greater uncertainties associated with steeply rising resistive losses, exceeding \qty{5}{\percent}, in the metal of the antenna, mesh and ground. Therefore, only data in the range \qtyrange{60}{350}{\mega\hertz} are used for accurate comparison of radio sky models with the measurements. Extended Data Fig.~\ref{fig:receiver_small_box} shows the antenna up close.  There are ten solid and ten wire dipoles in each polarization and the feed lines for these 20 dipoles consist of a pair of rectangular tubes conducting the radiation received in that polarization to the vertex.  The antenna is designed for \qty{50}{\ohm} single ended feeding.

Sky measurements were made with the antenna placed at the centre of mesh S16-4 at the southern extremity of the SKA-Low telescope, within Inyarrimanha Ilgari Bundara, the CSIRO Murchison Radio-astronomy Observatory in Australia. The specific location was \ang{-27.05934}, longitude \ang{116.479561} and height \qty{320.0}{\metre}. The quoted height is that of the ground mesh in geodetic coordinates with respect to the reference ellipsoid in the International Terrestrial Reference System (ITRS) frame.  This mesh was one in a tight cluster of six meshes that were unpopulated at the time of our measurement.  Each mesh will eventually host 256 SKALA4.1 antennas to form one SKA-Low station. 

The mesh is constructed from \qty{4}{\milli\metre} diameter wire welded together on a \qtyproduct{50 x 50}{\milli\metre} square grid. This yields a ground plane conductivity of \qty{0.2}{\mega\siemens\per\metre}, lower than might be expected for a metallic structure due to galvanisation of the mesh.  The mesh is made from hot-dipped galvanised steel, which is steel with a coating of zinc that is in the form of layered zinc-steel alloy that lowers the conductivity substantially \citep{CozzaConductivity}.

Simulations were made to examine for the effect of ground pickup from below and beyond the mesh.  Antenna patterns derived from electromagnetic simulation using soil below the mesh of relative permittivity $\epsilon_\text{r}=4$ and conductivity $\sigma_\text{soil}=\qty{0.7}{\milli\siemens\per\metre}$ showed no difference to the patterns used in the current work that ignored soil below the mesh.  This indicates that insufficient currents or sidelobe power reaches the lossy ground in the first place to be able to dissipate there.

The antenna was pointed at the zenith and its polarizations were oriented in the EW and NS directions.  Our measurements were made with the EW-oriented dipole arms with the unused NS-oriented arms terminated in \qty{50}{\ohm}, which is close to the antenna impedance over most of the band.  Supplementary Fig.~\ref{fig:Beam_fwhm} shows the principal half-power beamwidths versus frequency for the selected EW-oriented linear polarization state, as simulated in the H-plane (NS) and E-plane (EW).  As expected for EW-oriented dipoles, the beamwidth is larger in the NS H-plane. The mean half-power beamwidth over the operational frequency range is \ang{64} in the EW plane and \ang{82} in NS plane.  

Small glitches disrupt beamwidth smoothness over frequency.  These are due to spurious radiation of secondary dipole harmonics and have been confirmed in anechoic chamber measurements of the antenna \citep{2023RS007758}. The antenna patterns from the electromagnetic simulations shown here, capturing the sharp features, have been directly used in deriving our offset and scale factors to correct GSM2016.  Hence, this anomaly in antenna behaviour has been accounted for in our analysis.

Supplementary Fig.~\ref{fig:beam_slices} shows antenna pattern cuts at \qty{145}{\mega\hertz} for three fixed azimuth angles.  These show a wider beam with larger sidelobes at azimuth \ang{0} compared to azimuth \ang{90}.  The response to the horizon goes to zero at all azimuths.  Supplementary Fig.~\ref{fig:beam_section} shows beam footprints on the sky at the \qty{3}{\decibel} and \qty{10}{\decibel} gain levels.

The SKALA4.1 antenna usually incorporates a low-noise amplifier (LNA) at the apex of each of its two polarizations. We replaced these with modified end caps to use the antenna with the external GINAN receiver that incorporates its own LNA.  The feed lines corresponding to the unused NS polarization dipoles were terminated at the apex in a \qty{50}{\ohm} load enclosed in an end cap (Supplementary Fig.~\ref{fig:antenna_end_caps}a). A \qty{3.1}{m} low-loss FSJ1 HELIAX\textregistered \  coaxial cable was passed through from base to apex of one of the feed lines corresponding to the EW polarization used for measurements. At the apex the coaxial cable is connected to an SMB coaxial RF connector on a small printed circuit board that connects the cable ground with that of the feed leg hosting the cable and connects the cable core via a short \qty{50}{\ohm} transmission line to the opposite feed line. Supplementary Fig.~\ref{fig:antenna_end_caps}b shows the resulting connectorised end cap.  Both \qty{50}{\ohm} load end cap and connectorised end cap are modifications of the original SKALA4.1 LNA cap assembly shown in Supplementary Fig.~\ref{fig:antenna_end_caps}c.  

The GINAN receiver was split between a pair of enclosures, a small unit close to the antenna base (Extended Data Fig.~\ref{fig:receiver_small_box}) to which the \qty{3.1}{\metre} coaxial cable connects, and a larger unit placed \qty{19}{\metre} away in a small tent at the edge of the ground mesh (Fig.~\ref{fig:Boolardy_Deployment}).  The smaller and larger units are connected by a \qty{19}{\metre} low-loss LDF4-50A HELIAX\textregistered \ coaxial cable  to receive the radio frequency signal from the antenna. They are also connected by a second lower-quality coaxial cable for the purpose of sending control pulses from the larger box to an RF switch in the smaller box that controls whether the receiver is connected to a particular calibration load or the antenna.

\subsection{Wideband radiometric calibration}\label{sec:receiver_calibration}

As stated in Section~\ref{radiometer}, in each calibration cycle the switch cycles through all five RF loads and the antenna.  Measurements are made in each of these six switch positions with both the vector network analyser (VNA) and spectrometer to derive receiver and antenna calibrations.  The five RF loads are precision short, precision open, precision \qty{50}{\ohm}, a network with complex impedance, and a calibrated noise source that is temperature compensated.

Based on measurements within each calibration cycle, the error terms in the VNA are first solved for using reflection coefficient measurements of standard precision terminations. With those corrections applied the complex impedance of the antenna, receiver, complex load, and the internal calibrated noise source are measured.  With knowledge of these complex impedances and using measurements made by the spectrometer of the power spectra from the set of internal terminations, which includes a \qty{50}{\ohm} termination with temperature probe attached and the temperature-compensated excess noise source, a joint solution is derived for the noise waves and bandpass.  The noise waves are described by four real-valued parameters and the bandpass describes the passband shape and also provides a scaling of spectrometer counts to antenna temperature in kelvin. The noise waves and scale are referred to the input terminals of the GINAN receiver.  

The GINAN receiver performs this full internal calibration in cycles with a cadence of \qty{4}{\minute}. This time was set based on laboratory studies of the stability of the electronics. All calibration, including the four-parameter model for the noise waves, is derived on a per-channel basis and solved for independently in each frequency channel and in each calibration cycle.  The VNA provides \num{2001} independent measurements over \qtyrange{30}{350}{\mega\hertz}; the spectrometer provides \num{4001} measurements with \qty{100}{\kilo\hertz} resolution bandwidth over \qtyrange{30}{350}{\mega\hertz}.  Impedance computations at the spectral measurement points of the network analyser are interpolated to the channelisation frequencies of the spectrometer.  The calibration is applied to the spectrometer measurement made with the switch connected to the antenna plus its feed cable to derive calibrated sky spectra for each calibration cycle.  These calibrated sky spectra are of antenna temperature at the terminals of the GINAN receiver on an absolute noise-temperature scale in kelvin, corrected for receiver noise waves, receiver bandpass, and for any impedance mismatch between antenna and receiver at the terminals of the GINAN receiver.  The calibrated spectra have the native resolution and channel spacing of the spectrometer.  

\subsubsection{Calibration of reflections and receiver noise waves}
\label{sec:noisewavecal}

The switch acts as the main calibration reference plane, and calibration is performed on a per-cycle basis. Each cycle includes measurements of all switch positions using both the VNA and the spectrometer. The architecture of the GINAN receiver allows for direct VNA measurement of the complex reflection coefficient of the antenna plus feed cable relative to the receiver.  The VNA also directly measures the complex reflection coefficients of the noise source, \qty{50}{\ohm} load, and complex RF load relative to the receiver. The simplified noise wave model shown in Supplementary Fig.~\ref{fig:nw_terms} is adopted and the measured complex reflection coefficients along with the noise wave parameters provide accurate estimates of receiver noise contained in each of the spectrometer measurements made with the switch connected to one of the RF loads or antenna.  Additionally, the impedance of the aggregate antenna and cable, as well as the impedance of the receiver, are determined separately relative to \qty{50}{\ohm} reference allowing the determination of the power transfer function into the radiometer.

With the switch connected to one of the six terminations, the complex reflection coefficient $\Gamma_{1t}$  of the termination impedance relative to the receiver impedance is given by
\begin{equation} \label{eq:two_term}
\Gamma_{1t} = (\Gamma_m - \epsilon_{00})/g_{21},
\end{equation} 
where $\Gamma_m$ is the reflection coefficient as measured by the VNA $S_{21}$ measurement with ports labeled as per Extended Data Fig~\ref{fig:radiometer_concept}; and where $\epsilon_{00}$ and $g_{21}$ are the leakage  and gain, respectively, within the receiver architecture between the two ports of the VNA.  The leakage and gain are solved for, separately in each VNA frequency channel and in each calibration cycle, using VNA measurements of reflection coefficients with the switch connected to the precision open and short terminations. 

Additionally, using the VNA measurement of complex reflection coefficient in each switch setting, the complex reflection coefficient $\Gamma_{2t}$ of that termination relative to \qty{50}{\ohm} reference is given by
\begin{equation}\label{eq:three_term}
\Gamma_{2t} = \frac{(\Gamma_m - b)}{(a-c\Gamma_m)},
\end{equation} 
where the VNA calibration terms $a$, $b$ and $c$ are solved for using VNA measurements of reflection coefficients $\Gamma_{mo}$, $\Gamma_{ms}$, and $\Gamma_{ml}$ that are made with the switch connected to the precision open, precision short and precision \qty{50}{\ohm} terminations, respectively.  These VNA calibration terms are related to the leakage, gain, and an additional source match term $\epsilon_{11}$ by the relations: $a = g_{21}-\epsilon_{00}\epsilon_{11}$, $b = \epsilon_{00}$ and $c = -\epsilon_{11}$.

The complex impedance of the GINAN receiver relative to \qty{50}{\ohm} reference is given by
\begin{equation}\label{eq:receiver_impedance}
Z_r = (50 + j0) \frac{(\Gamma_{mo} - \Gamma_{ml})}{(\Gamma_{ml} - \Gamma_{ms})}.
\end{equation} 
Therefore, the power transfer to the GINAN receiver from a termination with noise temperature $T_t$ and impedance $Z_t$ is $T_t (1- \zeta_{t} \zeta^{*}_{t})$, where $\zeta_{t} = (Z_r - Z^{*}_t)/(Z_r + Z^{*}_t)$.

The spectrometer measurement contains the additive sum of (1) this noise temperature from the termination transferred to the GINAN receiver $T_t (1- \zeta_{t} \zeta^{*}_{t})$, and (2) receiver noise generated in the GINAN receiver, which includes emission from the LNA and other active devices in the RF signal path, Nyquist-Johnson (thermal) noise from the passive components such as the couplers, attenuators, cables and connectors, and the additive quantisation noise in the analogue-to-digital converters.  As shown in Supplementary Fig.~\ref{fig:nw_terms}, the total receiver noise is modelled as a pair of noise voltage waves \citep{1130490} referred to the switch plane: $N_1$ travelling upstream towards the terminations or antenna and $N_2$ travelling downstream in the direction of RF signal flow.  Referred to the switch plane, the total receiver noise voltage arriving at the spectrometer is
\begin{equation}
V_\text{rec} = \Gamma_{1t} N_{1} + N_{2}
\end{equation}
where $\Gamma_{1t}$ is the complex reflection coefficient of the termination with respect to the receiver. 

Thus, the total noise power $P_\text{sys}$ received by the spectrometer in units of spectrometer counts is
\begin{align}
P_\text{sys} = T_\text{sys} B_p =  \left\{ T_{t} (1 - \zeta_{t} \zeta^{*}_{t}) + N_{1}N^{*}_{1}\Gamma_{1t}\Gamma^{*}_{1t} + N_{2}N^{*}_{2} + 2\Re[N_{1}N^{*}_{2} \Gamma_{1t}] \right\} B_p, \label{eq:total_noise_power}
\end{align}
where $T_\text{sys}$ is the total noise temperature received by the spectrometer in kelvin units and $B_p$ is the GINAN receiver bandpass function representing the spectral gain of the receiver that converts system temperature to spectrometer counts. $B_p$ has units of spectrometer counts per kelvin noise temperature.

Calibrating for additive receiver noise and the bandpass function requires solving for the five calibration terms $N_{1}N^{*}_{1}$, $N_{2}N^{*}_{2}$, $\Re[N_{1}N^{*}_{2}]$, $\Im[N_{1}N^{*}_{2}]$ and $B_p$.  Spectrometer measurements of the total power with the switch connected to the precision open, precision short, network with complex impedance, precision \qty{50}{\ohm}, and calibrated noise source provide a set of five linear equations with these five calibration terms as variables.  This set is used by a chi-squared solver to simultaneously estimate the five calibration terms.  In-situ temperature logger recordings of the precision \qty{50}{\ohm} termination, and laboratory calibration of the noise temperature of the temperature-compensated noise source (see Section~\ref{sec:calib_scale}) are used in the solution.  The calibration terms are derived independently for each calibration cycle and for each spectrometer frequency channel.  The network with complex impedance, which is used as one of the calibration RF loads at the switch, was selected so that solutions for all five GINAN receiver calibration terms are possible without singularities over the entire operating frequency range. This required the network to be an inductance.

Finally, the calibrated noise temperature $T_\text{t}$ available at the end of the antenna feed cable (at the switch input) may be evaluated by inverting equation \eqref{eq:total_noise_power} for all spectra where the switch was connected to the antenna.  Substituting in noise wave terms, bandpass, complex reflection coefficient of the antenna (including its cable) relative to the GINAN receiver, and the power transfer function $(1 - \zeta_{t} \zeta^{*}_{t})$ yields antenna temperature in kelvin from any given spectrometer noise power spectrum $P_\text{sys}$.  It is a unique aspect of the GINAN receiver that it can evaluate and apply all of these calibration terms every \qty{4}{\minute}, in situ while connected to the antenna.

\subsubsection{Calibration of the temperature scale}
\label{sec:calib_scale}

The absolute calibration scale is set by the spectral noise powers from the internal \qty{50}{\ohm} termination, whose temperature is monitored during the observations, and the temperature-compensated internal noise source.  Laboratory calibration is performed to accurately determine the noise temperature of the internal noise source and to model for any offset between the noise temperature of the internal \qty{50}{\ohm} termination and its physical temperature as given by a temperature probe affixed to that termination.  

In the laboratory, an external \qty{50}{\ohm} termination with a temperature probe affixed was placed in an insulated  water bath and connected at the input terminals of the GINAN receiver in place of the antenna.  With the GINAN receiver cycling through its calibration and acquisition of spectra, the bath is first maintained at a stable low temperature with ice, then replaced with boiling water and allowed to cool slowly to ambient temperature over hours.  With the noise temperature of the internal noise source assigned a value corresponding to its manufacturer's datasheet, all receiver calibrations were derived and applied. The calibration includes corrections for the power transfer from the external \qty{50}{\ohm} termination to the GINAN receiver, and for the receiver noise waves that are conditioned by the voltage reflection coefficient of this external \qty{50}{\ohm} termination impedance relative to that of the GINAN receiver.  The estimated noise temperature of the external \qty{50}{\ohm} termination that is immersed in the bath is then compared with the physical temperature reading of the probe that is affixed to that termination.  The probe temperature varies from close to \qty{0}{\degreeCelsius} to somewhat above \qty{70}{\degreeCelsius} and linear fits to the noise temperature versus probe temperature yield a slope and offset at each frequency bin.  

The factor by which the slope of the fit departs from unity is the fractional error in the value assigned to the noise source. The offset in the intercept from origin is the difference between the noise temperature of the internal \qty{50}{\ohm} termination and the reading provided by the attached temperature probe. Using the slopes of the linear fits over frequency we have modelled the profile of the noise source as shown in Supplementary Fig.~\ref{fig:Noise_source_model}. Using the offsets from the origin in the linear fits we have modelled the noise temperature of the internal termination in terms of the temperature probes affixed to the termination and within the GINAN receiver enclosure.

Adopting this laboratory calibration for the noise source and internal termination and hence for the temperature scale, the slope and offset in the linear fits to the noise temperature versus probe temperature are shown in Supplementary Fig.~\ref{fig:Slope_offset} over the \qtyrange{60}{350}{\mega\hertz} band. The fits were made at the native spectrometer channel spacing of \qty{80}{\kilo\hertz}. The slopes and offset values are also displayed after Hanning smoothing with a \qty{10}{\mega\hertz} window, where it is seen that for measurements made with such bandwidths the calibration has uncertainty with RMS value of \qty{0.5}{\percent} in scale and \qty{1.5}{\kelvin} in offset. 

\subsubsection{Calibration of the antenna cable}\label{sec:calib_cable}

The terminals of the SKALA4.1 antenna are located at its vertex.  A customised end cap (Supplementary Fig.~\ref{fig:antenna_end_caps}b) allows the antenna terminals to be connected by a low-loss coaxial cable to the GINAN receiver's RF switch, located in a small enclosure at the base of the antenna (Extended Data Fig.~\ref{fig:receiver_small_box}). Since the switch serves as the calibration reference plane, the cable characteristics need to be determined and the data corrected for the multiplicative cable transfer function and additive thermal noise. The \qty{3.1}{m} cable was positioned inside the antenna strut with its top end disconnected at the antenna vertex, and a temperature probe was added midway down its length. A precision \qty{50}{\ohm} termination was connected at the top end of the cable in place of the antenna and the GINAN radiometer was cycled through its measurements and calibration steps. This was repeated with the termination replaced by a precision open circuit and then a precision short.  The cable was at ambient temperature during this calibration measurement. The GINAN receiver calibration process provided accurate measurements of the impedance and thermal emission spectrum of the cable with these terminations. These measured quantities are related to the cable attenuation, physical temperature, velocity factor and characteristic impedance and hence were used to solve for these properties.

The noise-temperature spectrum emerging from the cable may be considered to be composed of three parts: (1) thermal emission generated in the cable resistance and directed downstream towards the receiver; (2) thermal emission generated in the cable resistance that is directed upstream away from the receiver and then reflected back by the impedance mismatch between the cable and the \qty{50}{\ohm}, open or short termination; and (3) interference between the coherent parts of the random voltages corresponding to these two noise components.  These receiver noise components may be modelled as noise waves \citep{1130490}.  The models for cable impedance and cable emission spectra are fitted to the measurements made by the VNA and spectrometer and cable properties derived. Supplementary Fig.~\ref{fig:Cable} shows the solutions derived from these measurements, for cable attenuation, velocity factor, and complex cable characteristic impedance.

Using these solutions for the cable properties along with measurements of the physical temperatures of the cable and the \qty{50}{\ohm} termination at the cable end that were made using temperature probes, we predict the noise-temperature spectra at the GINAN receiver terminals owing to thermal emission of the cable.  These predictions are compared with measurements in Supplementary Fig.~\ref{fig:Cable_emission}.  For a reasonably well matched antenna, we expect the errors in our modelling of the characteristics of the \qty{3.1}{\metre} cable, along with in situ measurements of its temperature, to enable correction for cable additive thermal emission with RMS error less than \qty{1}{\kelvin}.  The cable attenuation is less than $10\%$, and the accuracy of the measurement of the cable attenuation by the VNA embedded in the receiver is about $1\%$. Therefore, the error in cable attenuation results in scale calibration error less than about $0.1\%$. 

Note that the GINAN receiver is designed to connect directly to an antenna's terminals. Avoiding a cable between antenna and receiver results in lower loss and smoother bandpass. This simplifies the calibration required to detect the very weak signals that GINAN targets, like the all-sky signal of the redshifted \qty{21}{cm} line during cosmic dawn.  There is no need to additionally calibrate and model attenuation and emission from the antenna feed cable when the GINAN receiver is directly connected to the terminals of an antenna.  Direct connection removes one of the few remaining manual calibration steps in the field (attaching calibration loads to the antenna end of the feed cable), removes the need to measure the temperature of the feed cable, and will reduce measurement uncertainty.

For the absolute temperature measurement of the radio sky that we have made using the GINAN receiver and SKALA4.1 antenna, the cable is inevitable, and hence needs to be accounted for.  The noise waves that our derived model predicts for the cable, when used to connect the SKALA4.1 antenna to the GINAN receiver, are shown in Supplementary Fig.~\ref{fig:cable_power}.  The calibrated noise temperature available at the base of the antenna cable, following the calibration steps discussed above in Section~\ref{sec:noisewavecal}, is corrected for these noise waves that our model predicts for the cable and for the measured attenuation in the antenna cable.  Additionally, a correction is made for the power transfer from the SKALA4.1 antenna to the antenna cable, based on derivations of the impedances of the antenna at its vertex, and of the GINAN receiver as referenced at the top end of the antenna cable at the antenna vertex.  This yields calibrated antenna temperature spectra for each calibration cycle. 

\subsubsection{Calibration validation}

The measurement data of the spectrometer were corrected to derive accurate radio sky brightness spectra in the form of antenna temperature spectra. The various stages of correction are illustrated in Supplementary Fig.~\ref{fig:waterfall_brightness}. Internal radiometer corrections were applied first: the receiver system multiplicative bandpass was calibrated for, then the spectrum was corrected for receiver additive noise. Next the attenuation and additive thermal noise from the cable connecting the antenna to receiver were calibrated for, and the transfer function through antenna to the cable (including antenna-receiver impedance mismatch) was calibrated out.  Lastly, correction was made for antenna radiation efficiency, which includes the resistive loss in the antenna and \qty{40}{\metre} ground mesh. Each calibration step resulted in smoother measured spectra that were closer to the GSM2016 predicted antenna temperature in both amplitude and spectral index.  

The precision of the end-to-end calibration algorithm and implementation was tested by using an antenna emulator. This consisted of a poorly matched RLC circuit with a real impedance of \qty{25}{\ohm}. Once the calibration was applied, this produced a flat spectral response at \qty{292}{\kelvin} corresponding to ambient temperature. The standard deviation of the spectrum over the entire band was approximately \qty{400}{\milli\kelvin}. This provides confidence that the calibration scheme is removing both the multiplicative and additive error while also correctly scaling the result. Additionally, we verified the power transfer (impedance mismatch) correction made by the GINAN receiver by extracting the complex impedance of the antenna from the receiver calibration parameters.  This agreed well with prediction from electromagnetic simulation of the antenna and ground mesh. 

\subsection{Systematic uncertainties}\label{sec:uncertainty}

The total systematic uncertainty in our measurement of the sky radio brightness is the quadrature sum of errors arising from absolute receiver calibration uncertainty; antenna cable attenuation and emission uncertainties; antenna pattern uncertainty; and the uncertainty arising from our use of a single linear polarization for measurement of radio sky brightness, which may be polarized.  

\subsubsection{Absolute calibration uncertainty}\label{sec:uncertainty_abs_cal}
The absolute calibration of the GINAN receiver measurement is set by laboratory calibration of the internal noise source as described in Section \ref{sec:calib_scale}.  The dynamic self-calibration procedures used in the field measurements were all used in the laboratory calibration.  This included calibrations for bandpass profile and gain, receiver noise waves, and impedance mismatches.  The residuals of the linear fits to calibrated noise temperature versus probe temperature in Section \ref{sec:calib_scale} demonstrate that the absolute calibration of the temperature scale has uncertainty with RMS value of \qty{0.5}{\percent} in scale and \qty{1.5}{\kelvin} in offset. 

\subsubsection{Cable attenuation and emission uncertainties}
The \qty{3.1}{\metre} cable connecting the antenna to the receiver attenuates the received signal and also adds its own thermal noise to the sky signal received.  The measurement data are calibrated for the cable based on accurate laboratory measurements of the cable properties that were made using the GINAN receiver and in-situ measurements of the impedance mismatches in the signal propagation from antenna via this cable to the receiver.  These measurements and derived corrections are discussed in Section \ref{sec:calib_cable} and the error in cable thermal emission, which appears as an additive to be subtracted from measurement data, is estimated to have RMS value less than \qty{1}{\kelvin}.  The error in cable attenuation results in scale calibration error less than about \qty{0.1}{\percent}.

\subsubsection{Antenna pattern uncertainty}
 The antenna pattern computed from electromagnetic (EM) simulations is convolved with the model sky to give an expectation for the calibrated measurement.  Errors in the EM simulation model for the pattern translate to errors in expectation for the measurement, given any global sky model.  Consequently, comparisons between measured and expected sky brightness yield erroneous corrections for the model sky.  The errors in EM simulations are usually quantified by comparing results of simulations made using different packages and methods.  The isolated SKALA4.1 antenna on ground plane was modelled \citep{9107113} using \texttt{FEKO} and, separately, \texttt{CST}, \texttt{MWS} and the directivities at zenith and off zenith agreed well.  Comparisons have also been made \citep{2158_1349732} for this antenna between \texttt{FEKO} and \texttt{Galileo}.  
 Differences in amplitude gain show that errors in individual patterns are not random over azimuth/elevation, but systematic, and hence would not be expected to diminish in the averaging over the pattern.  Second, the error patterns are substantially different at different frequencies.  Lastly, the magnitude of the error in power pattern is less than \qty{1}{\percent}.  It follows that the forward modelling of global sky model brightness to the measurement may result in less than \qty{1}{\percent} error at individual frequencies, with uncorrelated errors across the band. 

\subsubsection{Uncertainty from use of a single linear polarization mode}
Our measurement of the radio sky intensity is made using a single linear polarization mode; therefore, sky polarization is expected to manifest in confusing spectral structure.  Simulations \citep{2019MNRAS.489.4007S} of the polarized radio sky and Faraday depth at long wavelengths show that RMS values of spectral confusion arising from measurements of the total intensity using single-polarization radiometers would be in the range of \qtyrange{150}{200}{\milli\kelvin}, with perhaps an extended tail up to \qty{400}{\milli\kelvin}, when averaged over the wide beams of broadband dipole-like antennas.  Hence errors in measurement of radio sky brightness in our observing band, using a single polarization mode, are expected to have RMS values less than \qty{0.4}{\kelvin} owing to polarization of the brightness.     

\subsubsection{Total measurement uncertainty}
Adding the above four sources of error in quadrature yields the RMS values of uncertainty in our measurement of the radio sky brightness temperature. We display this versus frequency as the dotted line in Fig.~\ref{fig:RMS_errors}.  At \qty{150}{\mega\hertz}, this RMS value of measurement uncertainty is \qty{1.22}{\percent}.

\subsubsection{Uncertainty in the derived corrections to GSM2016}
Any global sky model, including the GSM2016 for which we have derived corrections, may not only have errors in the form of a global offset and scale factor, but may, in addition, have systematic errors that vary over the sky.  Such errors would manifest in residual differences between the measurements and predictions for these measurements that are made after correcting the global sky model for best fit offset and scale factor.  The dash-dot line in Fig.~\ref{fig:RMS_errors} shows the RMS value of the fractional difference between our measurements and our corrected GSM2016; at \qty{150}{\mega\hertz}, this RMS value is \qty{1.33}{\percent}. Such errors in sky maps will confuse the derivation of global offset and scale factor corrections. 

Given the calibration accuracy of the GINAN receiver and the quality of modelling of the SKALA4.1 antenna, it is not surprising that the errors in our derived corrections to GSM2016 are dominated by errors in the images that are used to construct the sky model: errors that are not representable as global offsets and scale factors.  Fig.~\ref{fig:RMS_errors} shows how we add, in quadrature, these errors arising from such inaccuracies in GSM2016 (dash-dot line) to the RMS values of uncertainty in our measurement of the radio sky brightness temperature (dotted line) to get the total uncertainty in the corrected GSM2016 (solid line).  At 150~MHz, the uncertainty in corrected GSM2016 is \qty{1.81}{\percent}.

\section*{Data availability}
Our measurements are made available in machine readable format \citep{McKay2026_RadioSkyBrightness} and may be used for setting the absolute scale for all-sky images constructed in the \qtyrange{60}{350}{\mega\hertz} band.
The primary data are measurements of antenna temperature in kelvin versus frequency and LST.  RFI is masked and corrections have been applied for receiver noise waves, bandpass, and impedance mismatch between antenna and receiver.  Frequency-dependent antenna radiation patterns are included in machine readable format for the purpose of relating antenna temperature to sky brightness. The offset and scale factor corrections derived specifically for GSM2016 are provided in equations \eqref{eq:offset} and \eqref{eq:scale}; these are only valid in the \qtyrange{60}{350}{\mega\hertz} frequency range of our measurement.


\section*{Acknowledgements} \label{sec:acknowledge}This scientific work uses data obtained from Inyarrimanha Ilgari Bundara, the CSIRO Murchison Radio-astronomy Observatory. We acknowledge the Wajarri Yamaji People as the Traditional Owners and native title holders of the Observatory site. Establishment of Inyarrimanha Ilgari Bundara and the CSIRO Murchison Radio-astronomy Observatory are initiatives of the Australian Government, with support from the Government of Western Australia and the Science and Industry Endowment Fund. The authors thank SKAO for the use of a SKALA4.1 antenna, an SKA-Low station ground mesh, and facilitating the deployment of the GINAN receiver with the antenna at Inyarrimanha Ilgari Bundara, the CSIRO Murchison Radio-astronomy Observatory.  The authors also thank CSIRO observatory and site entity staff for their support of the deployment.

\section*{Author Contributions}
L.~McKay, A.~Chippendale, R.~Subrahmanyan and R.~Ekers planned the work.  L.~McKay, R.~Subrahmanyan, A.~Chippendale and A.~Dunning invented the receiver architecture.  L.~McKay, R.~Subrahmanyan and A.~Chippendale built the receiver, refined it through laboratory experimentation and performed the field measurements at Inyarrimanha Ilgari Bundara, the CSIRO Murchison Radio-astronomy Observatory. R.~Subrahmanyan and L.~McKay processed the field measurements. P.~Bolli and G.~Kyriakou performed electromagnetic analysis of the SKALA4.1 antenna and the ground mesh. R.~Subrahmanyan, L.~McKay, A.~Chippendale and R.~Ekers interpreted the results.  R.~Subrahmanyan, A.~Chippendale and L.~McKay wrote the draft manuscript.  All authors reviewed and edited the manuscript.

\section*{Competing Interests}
We declare no competing interests.

\begin{appendices}

\section*{}
\begingroup
\let\origtheHfigure\theHfigure

\setcounter{figure}{0}

\renewcommand{\thefigure}{\arabic{figure}}

\renewcommand{\theHfigure}{ext.\arabic{figure}}

\renewcommand\figurename{Extended Data Fig.}
\clearpage
\begin{figure}[t]
\centering
\includegraphics[width=1\textwidth,angle=0,origin=c]{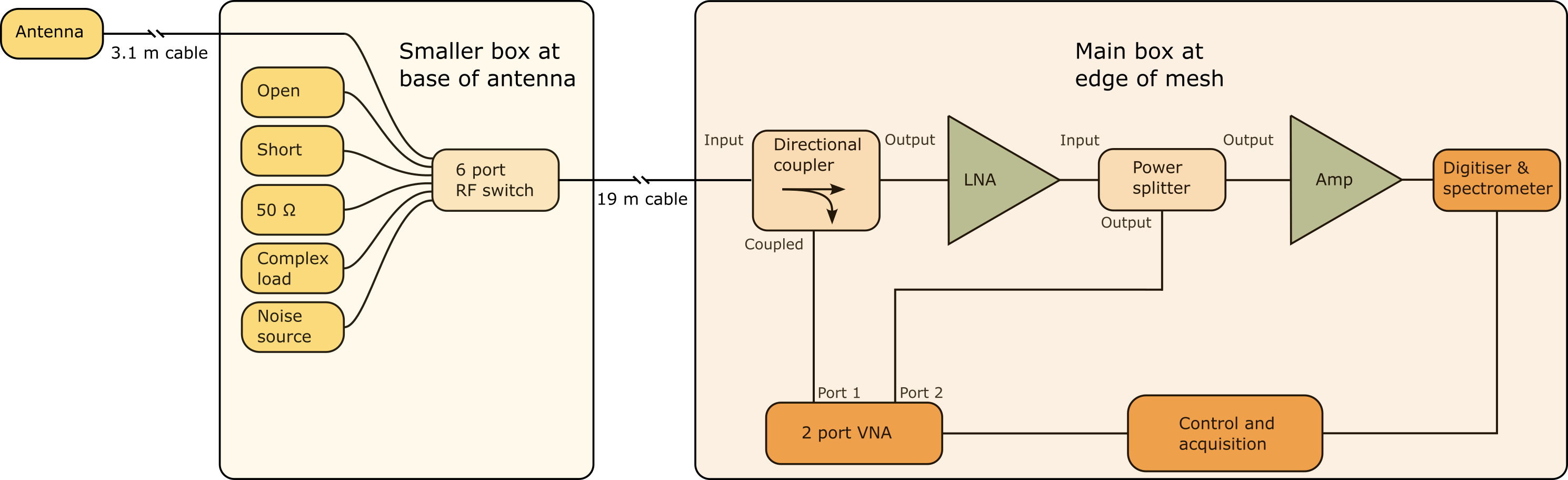}
\caption{\textbf{$|$ Novel GINAN radiometer architecture.}  The receiver cycles through four passive internal terminations, a noise source, and the antenna and measures each using both the vector network analyser (VNA) and spectrometer. These measurements enable calculation of the bandpass, receiver noise, and indeed all additive and multiplicative correction terms on a per-frequency-channel basis and with a cadence of minutes. This invention is protected by Australian patent application number 2025901969 filed on May 21, 2025 \citep{AU2025901969}. LNA: low-noise amplifier.
\label{fig:radiometer_concept}}
\end{figure}
\null

\clearpage
\begin{figure}[t!]
\centering
\includegraphics[width=0.9\textwidth,angle=0,origin=c]{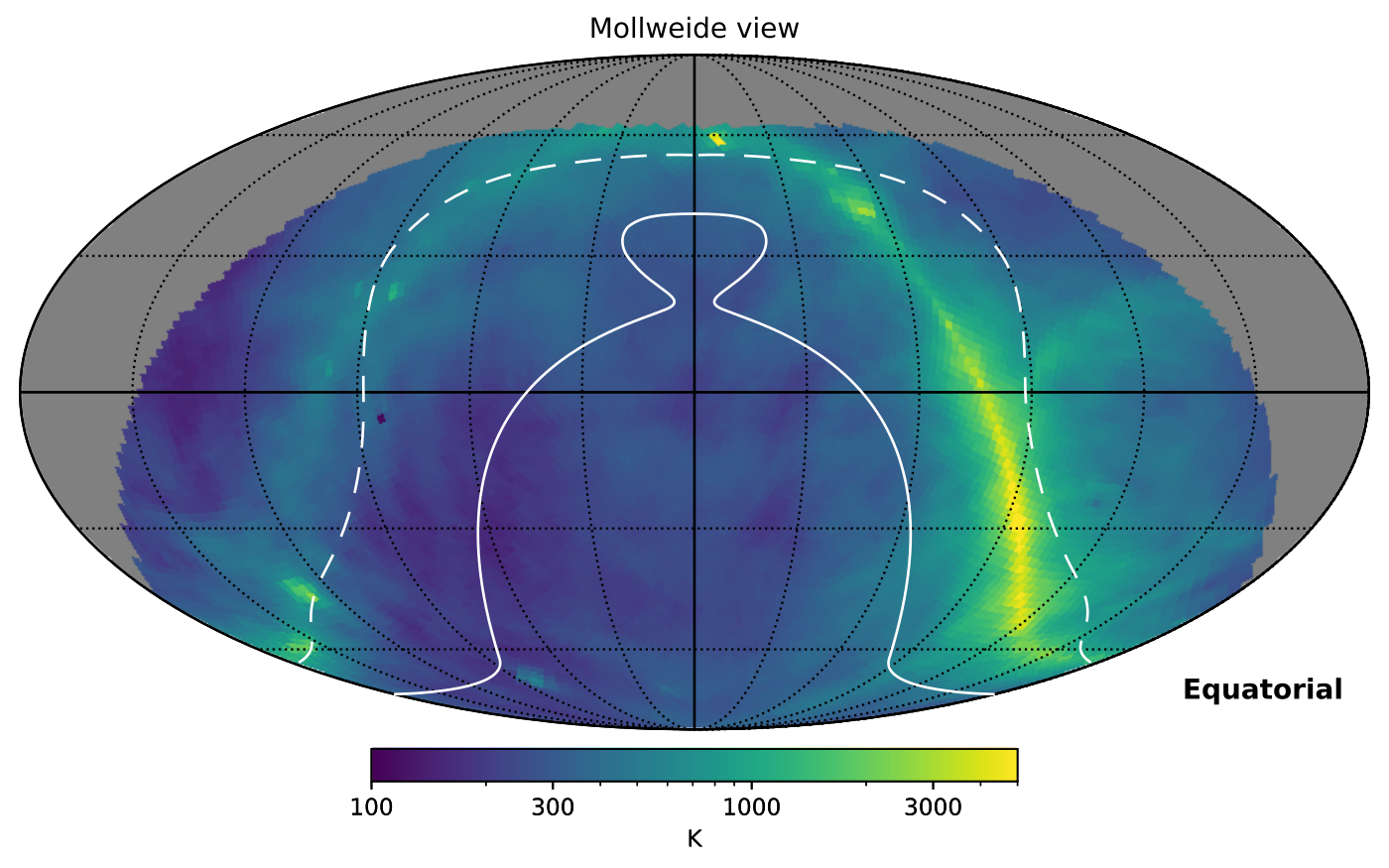}
\caption{\textbf{$|$ Sky coverage.} Sky coverage of our measurements shown by plotting the \qty{145}{\mega\hertz} radio sky \citep[GSM2016;][]{2017MNRAS.464.3486Z} in equatorial coordinates and truncating it (grey fill) where there is no coverage due to limited LST range or location of the antenna in the Southern Hemisphere.  The solid and dashed white contours respectively show where the effective gain of our measurements is 0.5 and 0.1.  The Mollweide projection of this map places RA \qty{12}{\hour} at the perimeter of the map and RA \qty{0}{\hour} on the vertical centre line with RA increasing to the left. The colour scale has an upper limit of \qty{5000}{\kelvin} with larger values plotted at this value.
\label{fig:sky_coverage}}
\end{figure}
\null

\clearpage
\begin{figure}[h]
\centering
\includegraphics[width=1.1\textwidth,angle=270,origin=c]{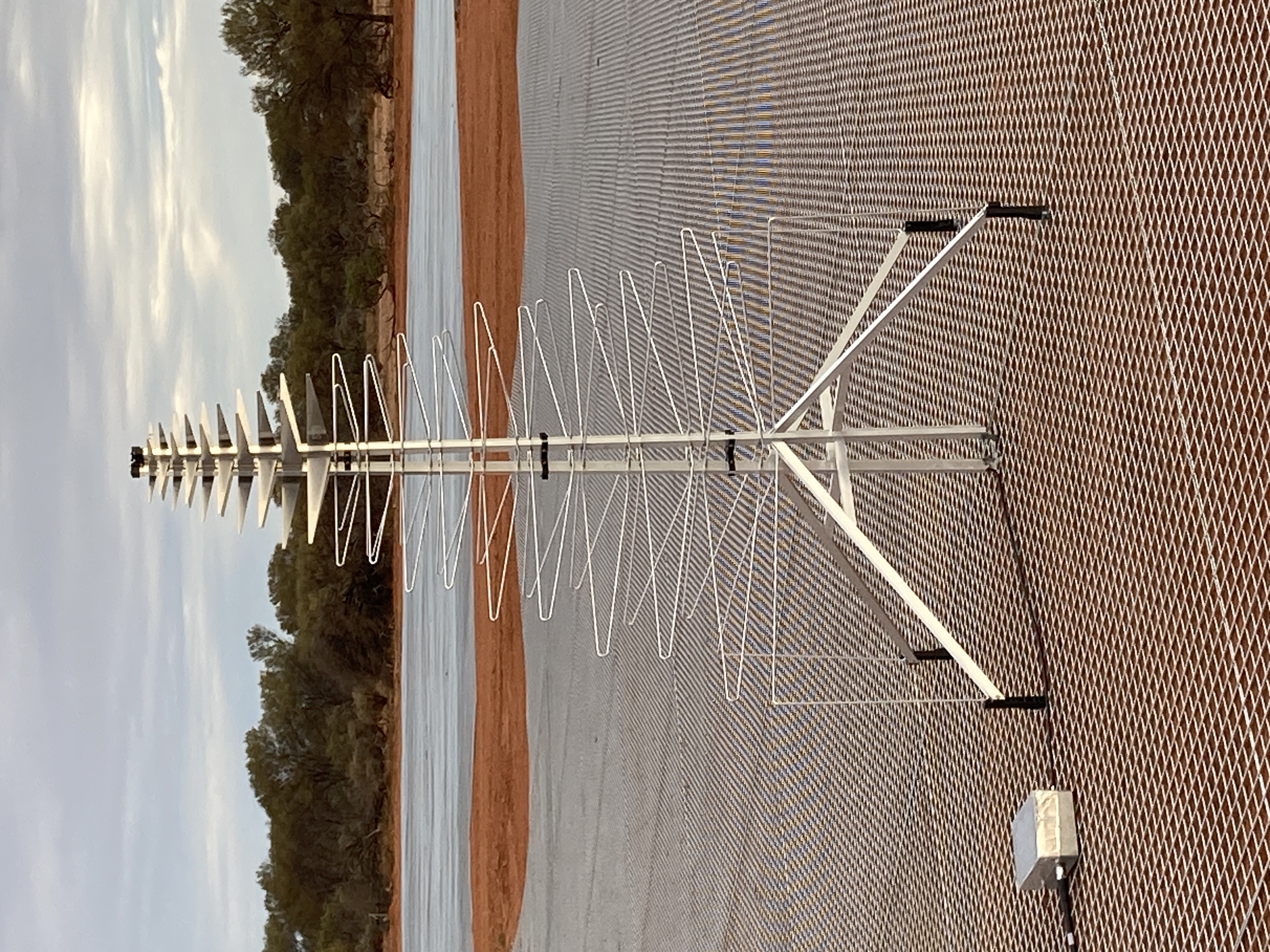}
\vspace{10pt}
\caption{\textbf{$|$ Antenna with smaller GINAN receiver box.}  The small rectangular shielded box at the base of the antenna contains a mechanical RF switch, a noise source, and four calibration loads.  The switch is cycled to connect the antenna, the noise source, then each of the four loads in turn to the main receiver box located near the edge of the ground mesh. 
\label{fig:receiver_small_box}}
\end{figure}
\null

\endgroup

\clearpage
\section*{}
\begingroup
\let\origtheHfigure\theHfigure

\setcounter{figure}{0}

\renewcommand{\thefigure}{\arabic{figure}}

\renewcommand{\theHfigure}{sup.\arabic{figure}}

\renewcommand\figurename{Supplementary Fig.}

\begin{figure}[h]
\centering
\includegraphics[trim=50 0 100 0,clip,width=1\textwidth,angle=0,origin=c]{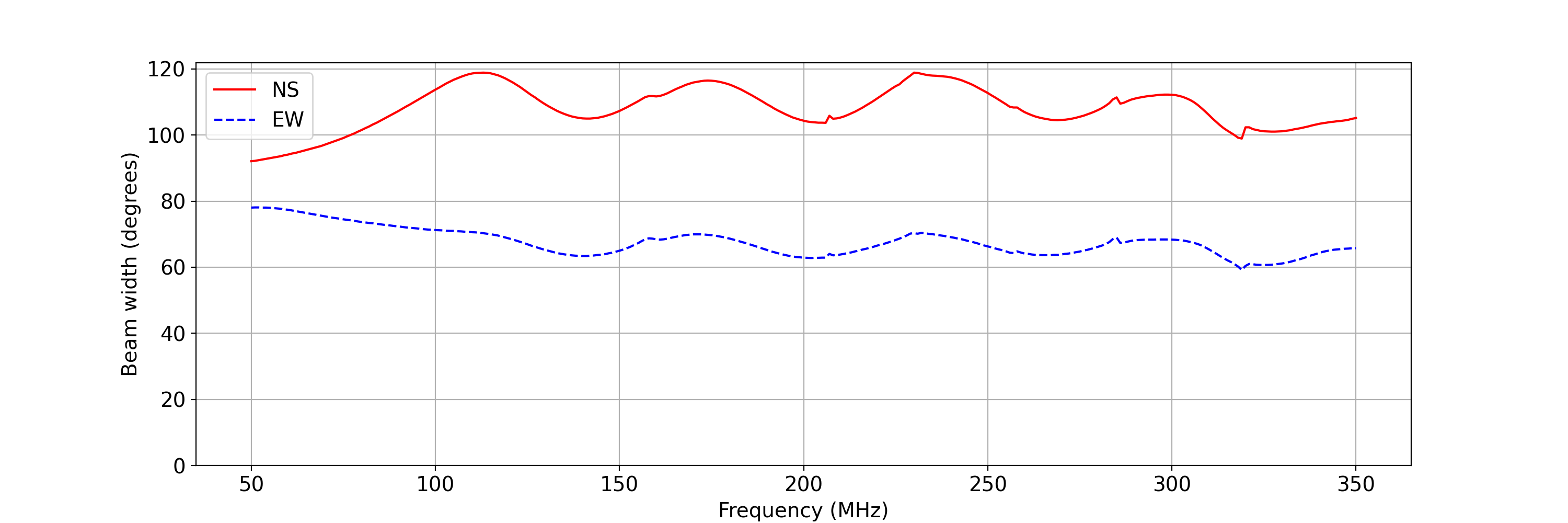}
\caption{\textbf{$|$ Antenna beamwidth.}  Half-power beamwidth versus frequency, for the SKALA4.1 antenna on conductive ground plane. Principal beamwidths along the H-plane (NS, solid line) and the E-plane (EW, dashed line) are shown, for the polarization state received by the EW-oriented dipole array of the antenna.  The small glitches disrupting beamwidth smoothness are due to spurious radiation of secondary dipole harmonics and have been confirmed in anechoic chamber measurements of the antenna \citep{2023RS007758}. The antenna patterns from the electromagnetic simulations shown here, capturing the sharp features, have been directly used in deriving our offset and scale factors to correct GSM2016.  Hence, this anomaly in antenna behaviour has been accounted for in our analysis.
\label{fig:Beam_fwhm}}
\end{figure}
\null
\clearpage

\begin{figure}[h]
\centering
\includegraphics[trim=113 0 120 0,clip,width=1\textwidth, angle=0, origin=c]{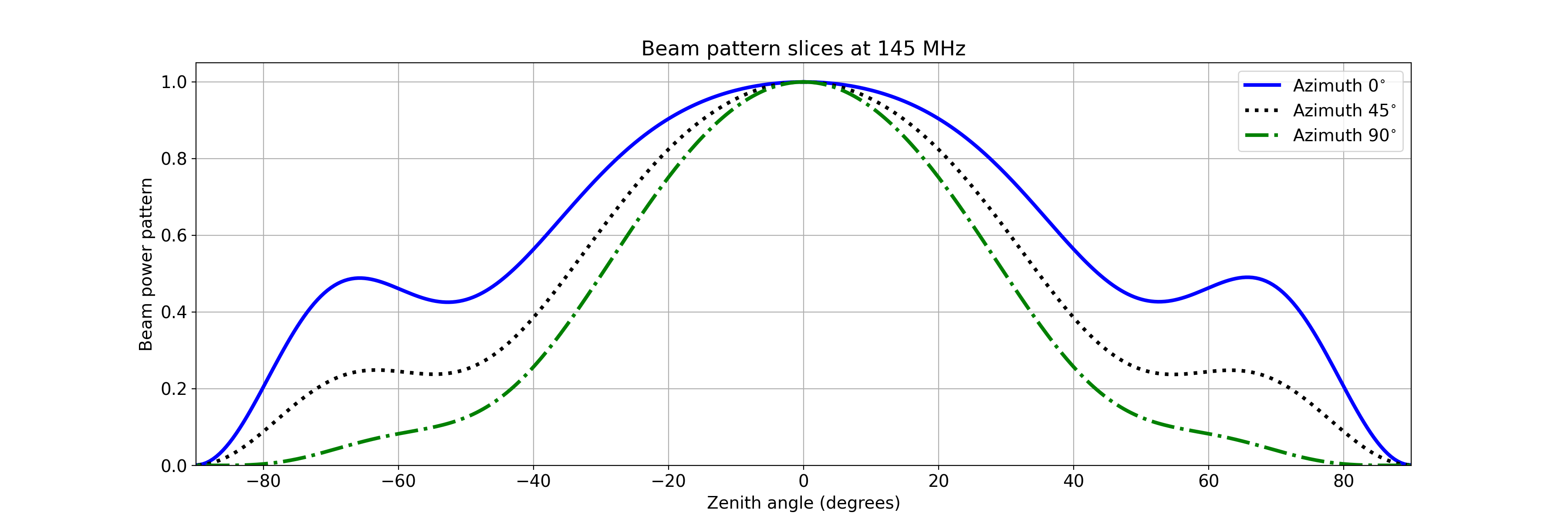}
\caption{ \textbf{$|$ Antenna beam cuts.} Cuts through the antenna beam at azimuth \qty{0}{\degree}, \qty{45}{\degree} and \qty{90}{\degree} at a typical frequency of \qty{145}{\mega\hertz}, which is close to the geometric mean of the band edge frequencies. The beam becomes wider and more distorted in going from azimuth \qtyrange{90}{0}{\degree}, with sidelobes emerging close to azimuth \qty{0}{\degree}.  The beam is directed at the zenith and the response to the horizon goes to zero at all azimuths. 
\label{fig:beam_slices}}
\end{figure}
\null

\clearpage
\begin{figure}[h]\centering
\includegraphics[width=1\textwidth, angle=0, origin=c]{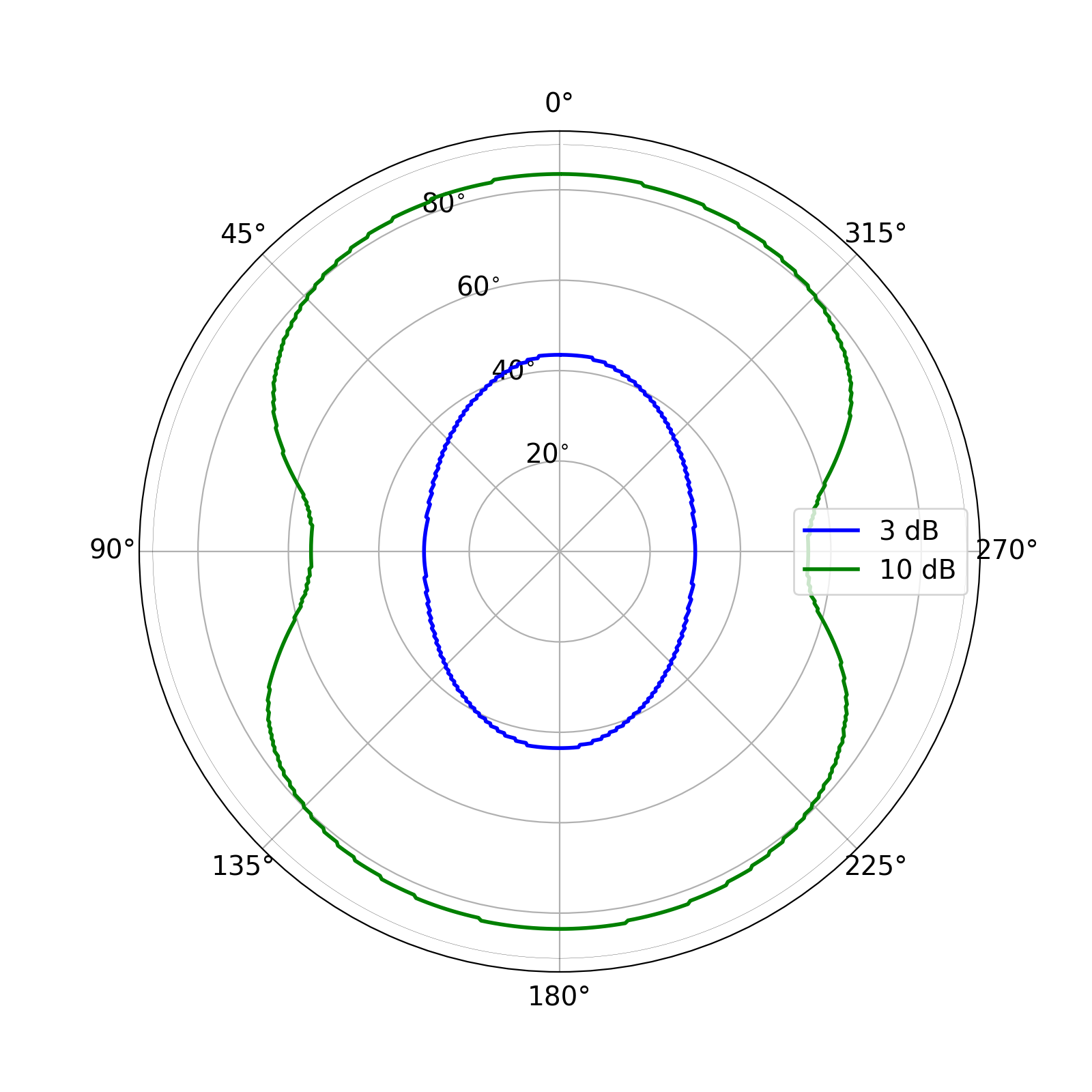}
\caption{ \textbf{$|$ Antenna beam footprints.}  Footprints of the beam pattern on the sky at \qty{3}{\decibel} and \qty{10}{\decibel} gain levels.  This plot shows the direction on the sky, in terms of azimuth and zenith angles, that the beam meets the specified gain level.  Zenith angle is  measured radially from the centre of the plot. The beam is directed at zenith and is wider in the NS plane than the EW plane.  For our measurements, the beam footprint was fixed in azimuth/elevation coordinates for all times and thereby all LSTs.
\label{fig:beam_section}}
\end{figure}
\null

\clearpage
\begin{figure}[h]
\centering
\includegraphics[width=1\textwidth,angle=0,origin=c]{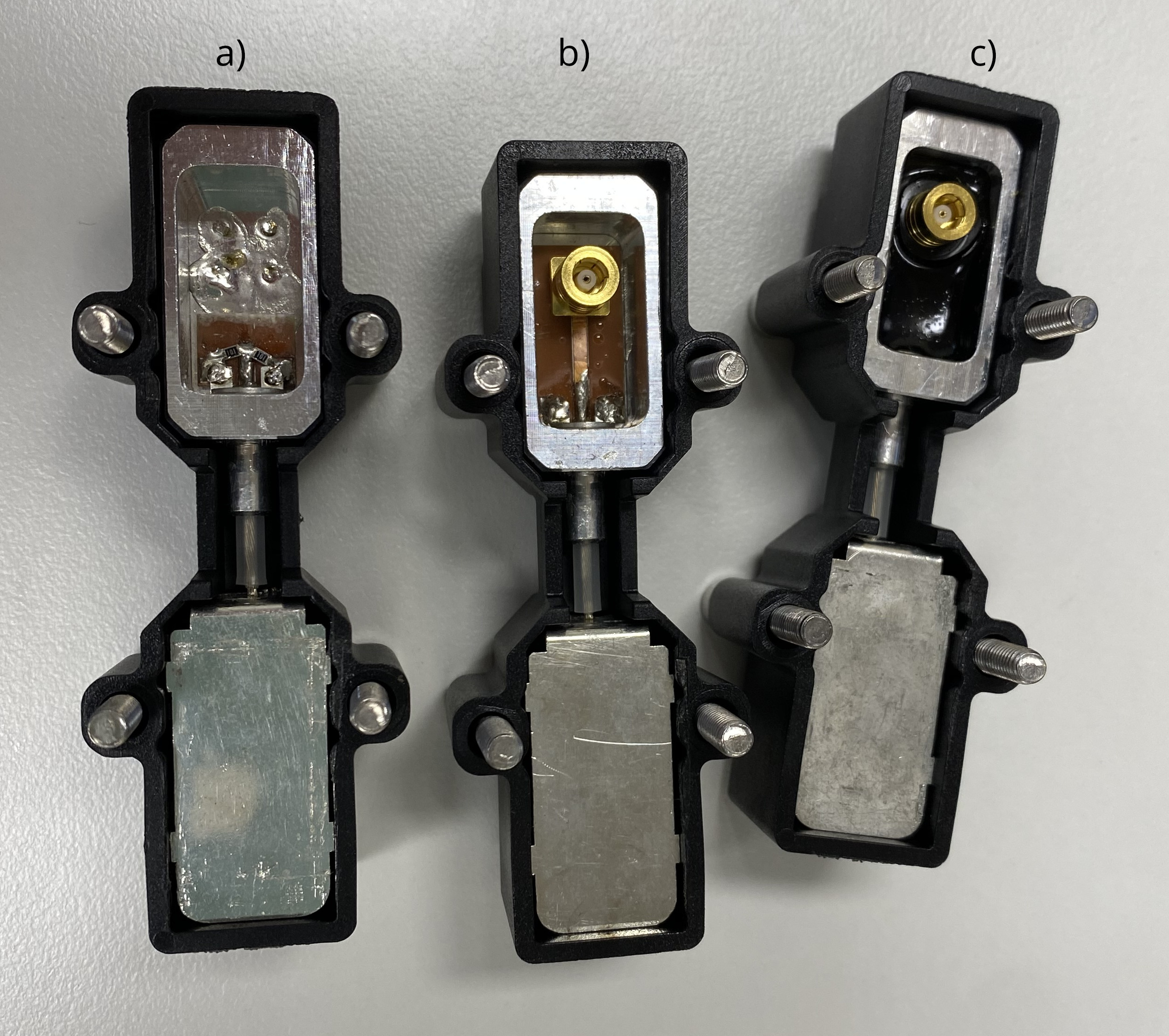}
\caption{\textbf{$|$ Antenna end caps.} \textbf{c}, The original SKALA4.1 antenna end cap with embedded LNA for installation at the vertex of each antenna polarization. \textbf{b} The modified end cap we used for measurements in this work, where the LNA is replaced with a direct microstrip connection to an SMB coaxial RF connector.  The antenna may then be connected via coaxial cable to an external receiver. \textbf{a}, Modified end cap that replaces the LNA with a \qty{50}{\ohm} load.  This was used to terminate the unused antenna polarization of the antenna for our measurements.  
\label{fig:antenna_end_caps}}
\end{figure}
\null

\clearpage
\begin{figure*}[h]
\centering
\includegraphics[width=0.7\textwidth]{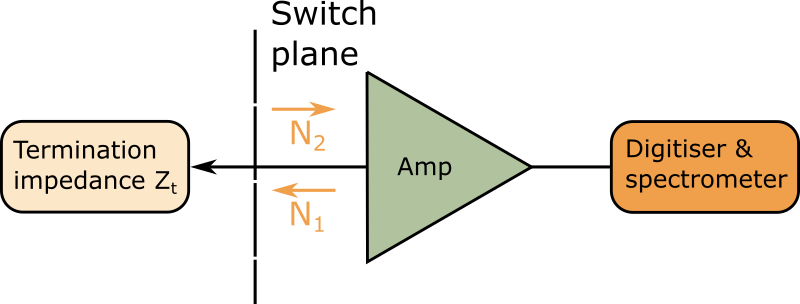}
\caption{\textbf{$|$ Noise wave model.} The noise wave terms $N_{1}$ and $N_{2}$ travel upstream and downstream respectively. The $N_{1}$ component reflects off of the terminating impedance before it is amplified by the RF link and recorded by the digital spectrometer. Noise contributions to measured spectra are therefore highly dependent on the receiver input impedance and its match to the impedance of the antenna or load connected to the receiver.
\label{fig:nw_terms}}
\end{figure*}
\null

\clearpage
\begin{figure}[t!]
\centering
\includegraphics[trim=70 0 90 0,clip,width=1\textwidth,angle=0,origin=c]{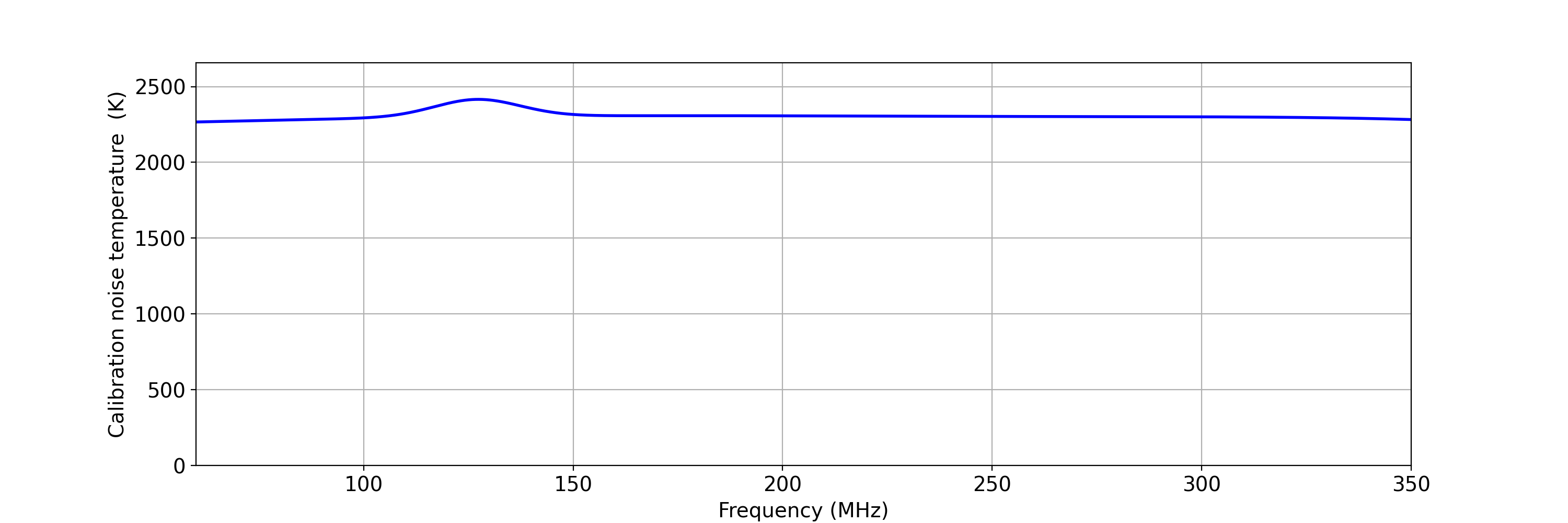}
\caption{\textbf{$|$ Laboratory calibration of noise source.} The model for the noise temperature of the internal calibration noise source derived from laboratory calibration.  The noise temperature of the internal source was calibrated via measurements of a precision \qty{50}{\ohm} termination placed at the input port of the GINAN receiver in place of the antenna, with the physical temperature of the termination varied by placing it in a bath with varying temperature that is known via an independent temperature probe.  The noise source couples into the receiver chain via a set of couplers and attenuators.  We attribute the small deviation observed in the \qtyrange{100}{150}{\mega\hertz} band to the bandshape of the noise coupling. 
\label{fig:Noise_source_model}}
\end{figure}
\null

\clearpage
\begin{figure}[t!]
\centering
\includegraphics[trim=30 0 10 0,clip,width=1\textwidth,angle=0,origin=c]{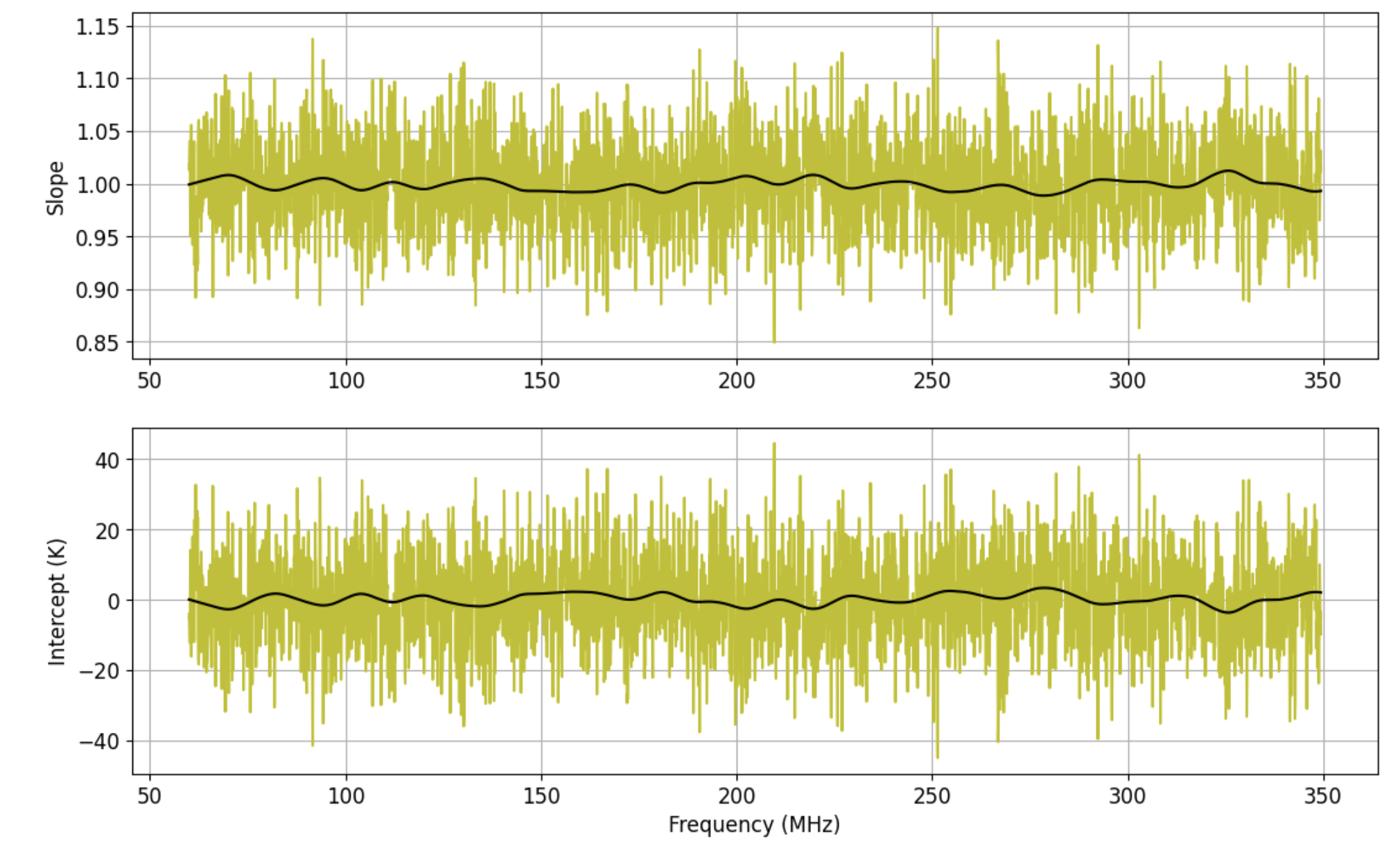}
\caption{\textbf{$|$ Verification of noise source calibration.} Fits to the scatter plots of the noise temperature of the \qty{50}{\ohm} termination in a thermal bath measured by the calibrated GINAN receiver, versus the physical temperature of the same termination measured by an independent temperature probe.  Slope and intercepts are shown in the two panels. Olive traces show the fit results per frequency channel and black traces show values averaged in \qty{10}{\mega\hertz} moving windows. For measurements made with such bandwidths the calibration has uncertainty with RMS value of \qty{0.5}{\percent} in scale and \qty{1.5}{\kelvin} in offset.
\label{fig:Slope_offset}}
\end{figure}
\null

\clearpage
\begin{figure}[t!]
\centering
\includegraphics[trim=10 0 10 0,clip,width=1\textwidth,angle=0,origin=c]{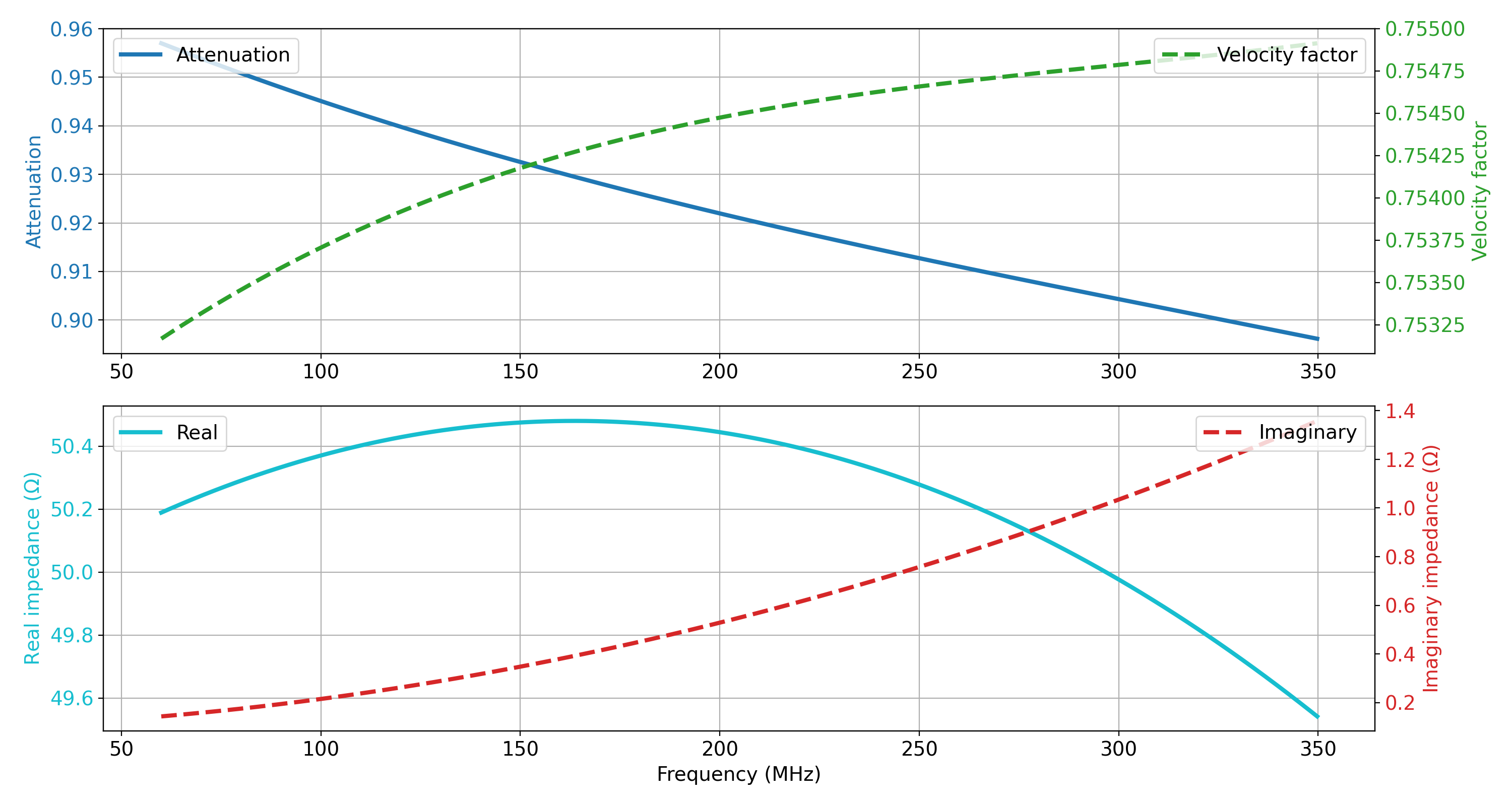}
\caption{\textbf{$|$ Antenna cable model parameters.} Characteristics of the FSJ1 HELIAX\textregistered \ coaxial cable connecting the antenna and GINAN receiver.  The upper panel displays the cable attenuation and velocity factor. The lower panel displays the real and imaginary components of the cable characteristic impedance, in units of ohms. The cable datasheet quotes a velocity factor of 0.82.
\label{fig:Cable}}
\end{figure} 
\null

\clearpage
\begin{figure}[t!]
\centering
\includegraphics[trim=80 0 100 0,clip,width=1\textwidth,angle=0,origin=c]{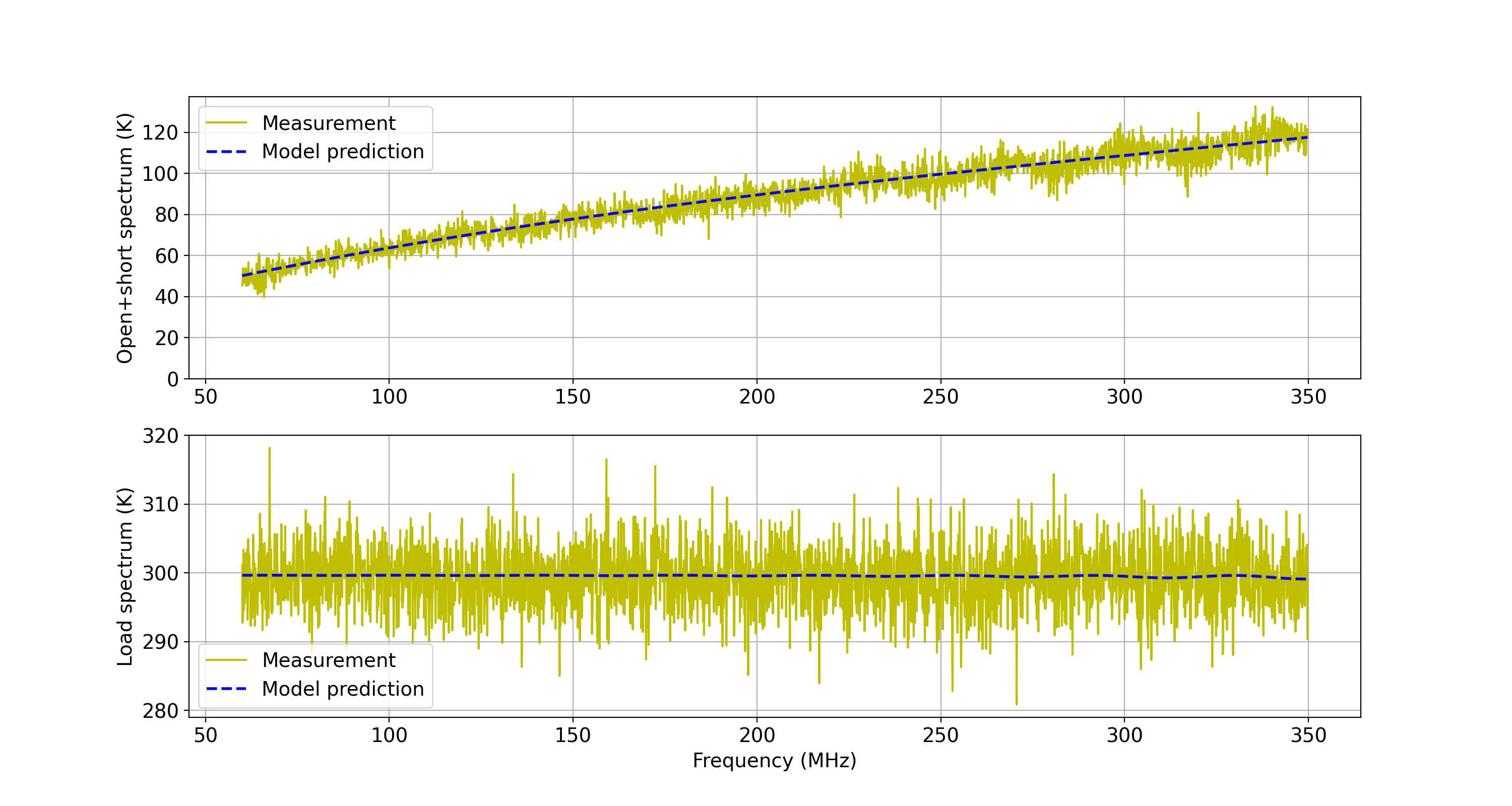}
\caption{\textbf{$|$ Antenna cable emission verification.} Predictions for cable thermal emission (dashed blue line), based on derived cable characteristics in Supplementary Fig.~\ref{fig:Cable}, compared with GINAN receiver measurements (solid olive line). The upper panel shows the sum of spectra acquired with the cable connected to the GINAN antenna terminals and with the far end of the cable terminated with a short and then an open circuit. The lower panel is for spectra acquired with a precision \qty{50}{\ohm} termination at the far end of the cable. 
\label{fig:Cable_emission}}
\end{figure}
\null

\clearpage
\begin{figure}[t!]
\centering
\includegraphics[trim=75 0 80 0,clip,width=1\textwidth,angle=0,origin=c]{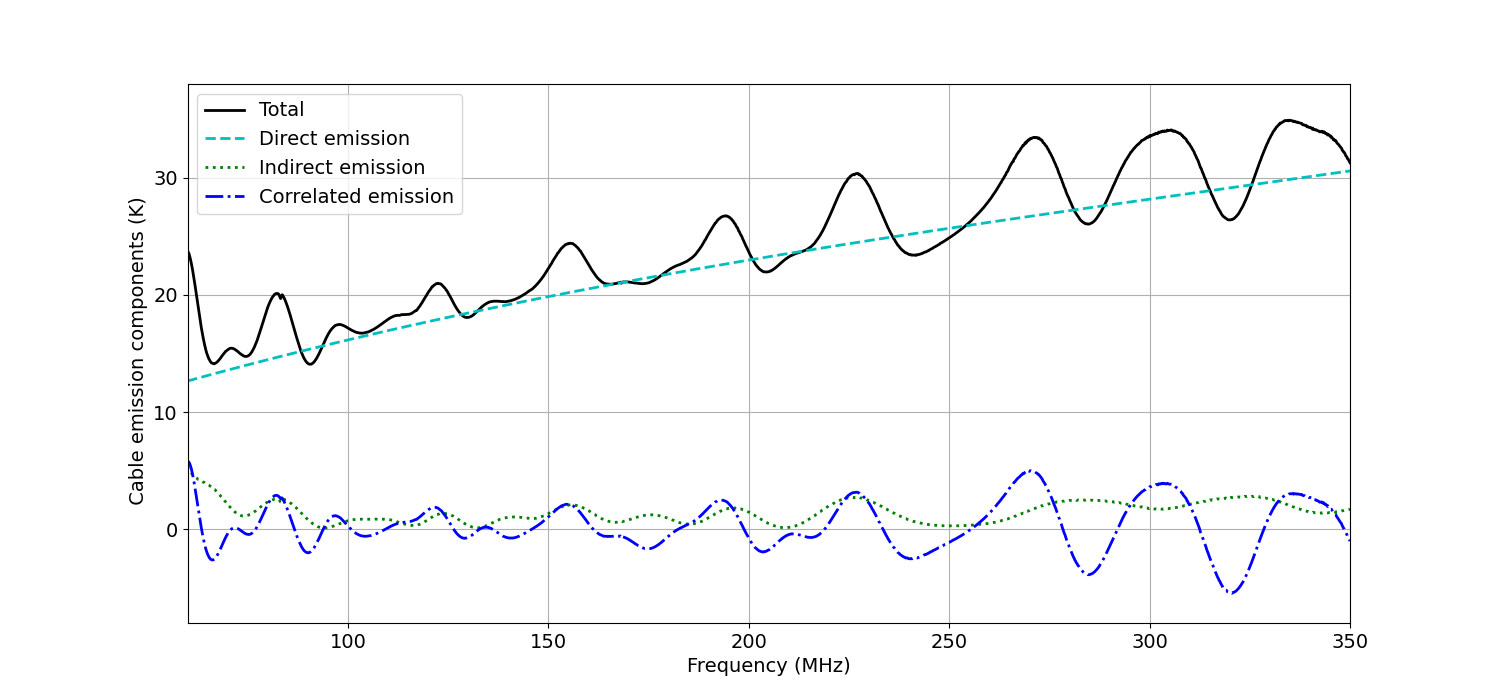}
\caption{\textbf{$|$ Measurement noise from antenna cable.}  Calculated noise wave components of the \qty{3.1}{\metre} cable, when connected between the SKALA4.1 antenna and GINAN receiver. Cyan (dashed) line: cable emission power directed towards the receiver. Green (dotted) line: cable emission power directed towards the antenna and arriving indirectly at the receiver, having reflected off the antenna. Blue (dash-dot) line: cross-correlation component that arises due to interference between correlated parts of the direct and reflected emission arriving at the receiver. Black (solid) line: sum total of these three contributions.   
\label{fig:cable_power}}
\end{figure}
\null

\begin{figure}[t!]
\centering
\includegraphics[trim=10 0 20 0,clip,width=1\textwidth,angle=0,origin=c]{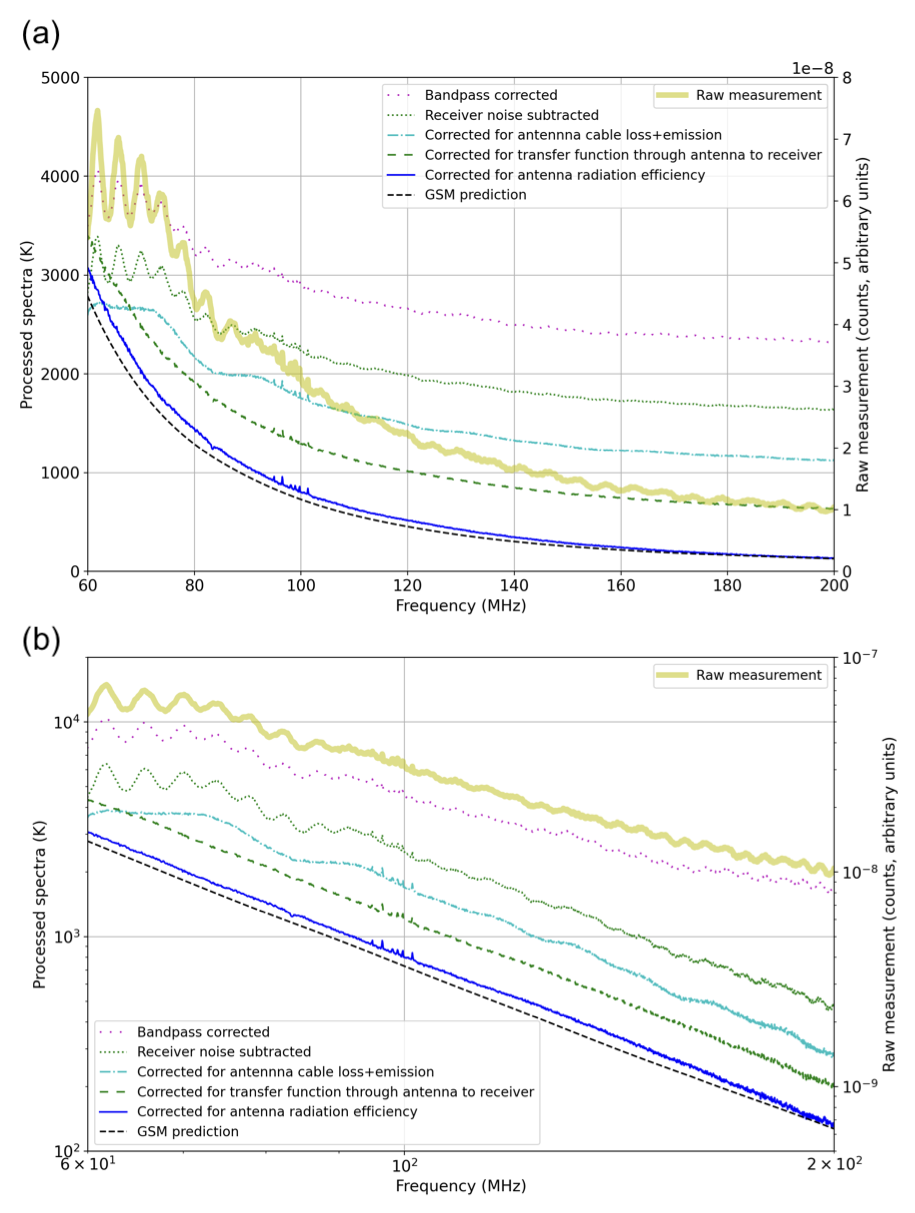}
\caption{\textbf{$|$ Breakdown of calibration steps.} Recorded spectra followed by different correction steps towards deriving the spectrum of the radio sky.  The frequency axis is limited to \qty{200}{\mega\hertz} for visual clarity. \textbf{a}, Plotted on a linear scale with an offset of \qty{500}{K} between subsequent traces. \textbf{b} Plotted on a log-log scale with a multiplicative scale factor of 1.5 between subsequent traces.
\label{fig:waterfall_brightness}}
\end{figure}
\null

\endgroup
\let\theHfigure\origtheHfigure
\end{appendices}
\end{document}